\documentclass[letterpaper,twocolumn,10pt]{article}
\usepackage{usenix-2020-09}

\usepackage{tikz}
\usepackage{amsmath}
\usepackage{xspace}
\usepackage{url}
\usepackage[most]{tcolorbox}

\usepackage{listings}

\usepackage{xcolor}
\usepackage{caption}

\definecolor{codegreen}{rgb}{0,0.6,0}
\definecolor{codegray}{rgb}{0.5,0.5,0.5}
\definecolor{codepurple}{rgb}{0.58,0,0.82}
\definecolor{backcolour}{rgb}{0.98,0.98,0.98}

\lstdefinestyle{mystyle}{
    backgroundcolor=\color{backcolour},   
    commentstyle=\color{codegreen},
    keywordstyle=\color{magenta},
    numberstyle=\tiny\color{codegray},
    stringstyle=\color{codepurple},
    basicstyle=\ttfamily\footnotesize,
    breakatwhitespace=false,         
    breaklines=true,                 
    captionpos=b,                    
    keepspaces=true,                 
    numbers=left,                    
    numbersep=5pt,                  
    showspaces=false,                
    showstringspaces=false,
    showtabs=false,                  
    tabsize=2,
    language=Python
}

\usepackage{xurl}
\usepackage{hyperref}
\usepackage{balance}
\usepackage[normalem]{ulem}
\usepackage{graphicx}
\usepackage{algorithm}
\usepackage{algpseudocode}
\usepackage{amsthm}
\usepackage{amsmath}
\usepackage{multirow}
\usepackage{url}
\usepackage{subcaption}
\usepackage{enumitem}
\usepackage{makecell}
\usepackage{booktabs} 
\usepackage{soul}
\usepackage{textcomp}
\usepackage{pifont}
\usepackage[T1]{fontenc}
\usepackage{adjustbox}
\usepackage{wrapfig}
\usepackage{microtype}
\usepackage{amssymb}
\usepackage{bbm}
\usepackage{tabularx}
\usepackage{bbding}
\usepackage{enumitem}

\usepackage[normalem]{ulem}

\newcommand{\yuheng}[1]{\textcolor{black}{#1}}
\newcommand{\shepherd}[1]{\textcolor{black}{#1}}
\newcommand{\areal}[1]{\textcolor{black}{#1}}
\usepackage[most]{tcolorbox}

\usepackage{tikz}

\newcommand\wcircle[1]{\raisebox{-0.5ex}{\Large\ding{\numexpr171+#1}}}

\theoremstyle{definition}

\newcommand{\PHB}[1]{\noindent\textbf{#1}}
\newcommand{\PHM}[1]{\vspace{.4em}\noindent\textbf{#1}} 
 
\newcommand{\SystemName}{\textsc{RollArt}\xspace}

\begin{document}
\pagenumbering{gobble}

\date{}



\title{\Large \bf \SystemName: Disaggregated Multi-Task Agentic RL Training at Scale}

\author{
    {\rm Wei Gao$^{\dagger*}$, Yuheng Zhao$^{\dagger*}$, Tianyuan Wu$^{\dagger*}$, Shaopan Xiong$^{\ddagger*}$, Weixun Wang$^{\ddagger*}$, Dakai An$^{\dagger}$,} \\
    {\rm Lunxi Cao$^{\dagger}$, Dilxat Muhtar$^{\ddagger}$, Zichen Liu$^{\ddagger}$, Haizhou Zhao$^{\ddagger}$, Ju Huang$^{\ddagger}$, Siran Yang$^{\ddagger}$,} \\
    {\rm Yongbin Li$^{\P}$, Wenbo Su$^{\ddagger}$, Jiamang Wang$^{\ddagger}$, Lin Qu$^{\ddagger}$, Bo Zheng$^{\ddagger}$, Wei Wang$^{\dagger}$}
    \bigskip \\
    {\em $^{\dagger}$HKUST \quad $^{\ddagger}$Alibaba Group \quad $^{\P}$Tongyi Lab, Alibaba}
}

\setlength{\floatsep}{3pt plus 2pt minus 1pt}
\setlength{\textfloatsep}{3pt plus 2pt minus 1pt}
\setlength{\intextsep}{3pt plus 2pt minus 1pt}
\setlength{\abovecaptionskip}{3pt plus 2pt minus 2pt}
\setlength{\belowcaptionskip}{3pt plus 2pt minus 2pt}
\setlength{\abovedisplayskip}{3pt plus 2pt minus 2pt}
\setlength{\belowdisplayskip}{3pt plus 2pt minus 2pt}

\maketitle
\begingroup
\renewcommand{\thefootnote}{*}%
\footnotetext{Wei Gao, Yuheng Zhao, Tianyuan Wu, Shaopan Xiong, and Weixun Wang contributed equally to this work.}%

\endgroup
\setcounter{footnote}{0}


\begin{abstract}

Agentic Reinforcement Learning (RL) trains LLMs through multi-turn interactions
with environments, producing workloads that mix compute-bound prefill,
bandwidth-bound decoding, CPU-heavy environment execution, and bursty reward
evaluation. Existing systems either colocate all stages on a single GPU cluster
or decouple them only at a coarse granularity, overlooking hardware
heterogeneity and incurring substantial synchronization overhead across stages.

We present \SystemName{}, a system for multi-task agentic RL on disaggregated
infrastructure. \SystemName{} maps each pipeline stage to best-fit hardware,
routing prefill-heavy tasks to compute-optimized GPUs, decode-heavy tasks to
bandwidth-optimized GPUs, and environments to CPU clusters. It decouples rollout
at the trajectory level, allowing generation, environment interaction, and
reward scoring to proceed independently, so that slow or failed environments
never block the others. \SystemName{} offloads stateless reward computation to
serverless infrastructure and overlaps rollout with training via
staleness-bounded asynchronous weight synchronization. Our results demonstrate
that \SystemName{} effectively improves training throughput and achieves
1.31--2.05\(\times\ \) training time reduction compared to various RL systems. 
\shepherd{We also evaluated \SystemName{} by training a hundreds-of-billions-parameter MoE model for Qoder product on an Alibaba cluster with above 3,000 GPUs, demonstrating its stability and scalability.}

\end{abstract}

\section{Introduction}
\label{sec:introduction}


Reinforcement Learning (RL) is advancing Large Language Models (LLMs) beyond
single-turn reasoning toward autonomous, long-horizon
decision-making~\cite{deepseekr1,seed-thinking,openaio4}. In this paradigm,
known as \emph{agentic RL}, an LLM interacts with external environments for
tool
use~\cite{hao2025exploringsuperiorfunctioncalls,wu2025agenticreasoningstreamlinedframework},
web
navigation~\cite{song2025r1searcherincentivizingsearchcapability,jin2025searchr1trainingllmsreason,jiang2025deepretrievalhackingrealsearch},
and computer
control~\cite{luo2025guir1generalistr1style,lu2025uir1enhancingefficientaction,liu2025infigui} over
multiple turns, learning to solve tasks through trial and error.

The agentic RL training pipeline operates as an iterative cycle of three stages:
\emph{rollout}, \emph{reward}, and \emph{training}. Unlike standard RL
post-training with single-turn responses, during rollout, the agent generates
actions and receives environment observations over multiple turns until a
termination condition is met, producing a \emph{long, multi-turn trajectory}.
Each trajectory is then evaluated in the reward stage, using rule-based
scripts~\cite{he2025deepmath,guo2024deepseekCoder}, code sandboxes, or separate
LLM
judges~\cite{son2024llmasajudgerewardmodel,zhong2024rlhfuseefficientrlhftraining}
to assign scalar reward signals. The training stage consumes scored trajectories
to update the agent's weights, which are synchronized back to the rollout
workers for the next iteration. 

As research labs scale agentic RL to larger models and harder
tasks~\cite{kimiteam2025kimik2openagentic,deepseek_v3_2_speciale}, a fundamental
resource-heterogeneity problem emerges \emph{across and within} these stages.
Within rollout alone, LLM generation alternates between compute-bound prefill
and bandwidth-bound decoding whose ratio depends on the task: on cost-equivalent
hardware, compute-optimized H800 GPUs cut prefill-heavy rollout time to
0.53$\times$ that of bandwidth-optimized H20 GPUs, while H20s cut decode-heavy
rollout time to 0.49$\times$--0.79$\times$ of H800s
(\S\ref{sec:motivation}). Concurrently, environments are \emph{stateful},
\emph{CPU-bound} processes whose latency is heavy-tailed due to host
contention, large variance in interaction turns, and environment failures. Reward workers are
\emph{stateless} and exhibit persistently low utilization---dropping to as
little as 7.4\% on dedicated GPUs---yet require elastic scaling when
trajectories complete. Finally, the training stage demands high-end GPUs with
fast interconnects. No single hardware type satisfies all stages.


A natural solution is to \emph{disaggregate} the pipeline and route
each stage to best-fit hardware. Yet, existing systems fall short.
\emph{Monolithic} frameworks such as veRL~\cite{verl,sheng2025hybridflow},
slime~\cite{slime_github}, and rLLM~\cite{rLLM} co-locate all stages on a single
GPU cluster and ignore heterogeneity entirely. \emph{Partially disaggregated}
systems such as AWorld~\cite{aworld} and DeepSWE~\cite{deepswe2025} offload
environments to Kubernetes~\cite{k8s} but still colocate the resource-heavy
rollout and training stages. Even state-of-the-art asynchronous systems address
only a subset of these heterogeneity dimensions: StreamRL~\cite{StreamRL},
AsyncFlow~\cite{asyncflow}, and SeamlessFlow~\cite{seamlessflow} split training
from rollout but treat rollout as a monolith, 
overlooking prefill/decode
divergence \emph{within} a task. 
AReaL~\cite{areal} introduces
unbounded-staleness asynchrony between training and generation, yet still
batches environment interaction and couples reward to generation GPUs.
Laminar~\cite{Laminar} schedules trajectories individually but does not exploit
within-rollout hardware affinity or offload stateless reward. As a result, no
existing system jointly addresses prefill/decode heterogeneity within generation, 
long-tail environment stragglers, stateless reward underutilization,
and cross-cluster bandwidth variability at production scale
(\S\ref{sec:motivation}).

To address these challenges, we present \SystemName{}, a distributed system
that maximizes throughput for multi-task agentic RL on disaggregated
infrastructure. The design of \SystemName{} is driven by four requirements derived
from empirical workload characterization (\S\ref{sec:motivation}):
\shepherd{\begin{enumerate}[label=\textbf{R\arabic*}, itemsep=2pt, topsep=2pt, parsep=0pt, partopsep=0pt]
    \item \textbf{Hardware-affinity workload mapping:} each pipeline stage, as well as sub-stages within rollout, should be bound to best-fit hardware so that compute-bound workloads land on compute-optimized GPUs and bandwidth-bound workloads on bandwidth-optimized GPUs. Because prefill/decode, CPU-bound, and stateless characteristics are stable per task domain in production workloads (\S\ref{sec:motivation} and \S\ref{sec:large-scale-analysis}), this binding can be driven by lightweight, domain-level annotations.
    \item \textbf{Trajectory-level asynchronous rollout:} environment interaction, LLM generation, and reward computation should be decoupled at the granularity of trajectories, so that environment stragglers never stall the pipeline.
    \item \textbf{Serverless offloading:} Stateless components, notably reward computation, should be offloaded to serverless infrastructure, eliminating dedicated hot standby GPUs while gaining autoscaling and fault tolerance.
    \item \textbf{Bounded-staleness asynchronous training:} training and rollout should execute concurrently on separate GPUs, overlapping weight synchronization with rollout. An asynchronous bound eliminates resource bubbles while preserving performance.
\end{enumerate}}


To meet these requirements, our system combines a declarative programming model
with a heterogeneity-aware distributed runtime. At job submission, users declare
hardware preferences for each stage through Python decorators. The runtime binds each stage to its best-fit resource
pool (\textbf{R1}): compute-optimized GPUs for training and prefill-heavy generation tasks, bandwidth-optimized
GPUs for decode-heavy generation tasks, CPU clusters for environments, and serverless endpoints for
reward. During rollout, the runtime drives each trajectory independently through
generation and environment interaction. A slow or failed environment never
blocks other trajectories, and reward scoring begins as soon as any trajectory
completes (\textbf{R2}). Completed trajectories are scored by stateless reward
functions on serverless infrastructure, granting elastic autoscaling without
dedicated GPU reservations (\textbf{R3}). While rollout continues, a separate
GPU cluster consumes scored trajectories for training. A bounded-staleness
weight synchronization protocol propagates updated weights back to the rollout
stage, overlapping training and cross-cluster transfer with ongoing generation
to eliminate resource bubbles (\textbf{R4}). Finally, cross-cutting
optimizations (\S\ref{subsec:cross-cutting-opt}) reduce cross-stage
data-movement cost, add resilience to environment failures, and support
optional prefill-decoding disaggregation within rollout.




We have implemented \SystemName{} in approximately 60k lines of Python code and
evaluated it by training Qwen3 models (8B--32B)~\cite{qwen-huggingface} on a
diverse mixture of agentic tasks across disaggregated H800 and H20 GPU clusters.
\SystemName{} achieves 2.05$\times$, 1.35$\times$, and \areal{1.31$\times$}, step-time reductions over a strengthened monolithic synchronous baseline, one-off baseline, and AReaL~\cite{areal}, respectively, and delivers
2.65--4.58$\times$ throughput over the synchronous baseline. Offloading
reward to a serverless cloud raises reward-GPU utilization from 6\% to 88\% and
halves per-step rollout time. We further quantify the overheads introduced by disaggregation, including cross-cluster weight synchronization via Mooncake~\cite{qin2024mooncake}, serverless reward I/O, and trajectory transfer, and show that they are outweighed by the resulting throughput gains (\S\ref{sec:eval}). 
\shepherd{To validate scalability and stability in a production setting, we have deployed \SystemName{} to
train a hundreds-of-billions-parameter Mixture-of-Experts (MoE) model for Qoder~\cite{qoder} product on a cluster of over 3,000 GPUs. This deployment runs
continuously for one week, confirming \SystemName{}'s ability to sustain high
throughput and robust fault tolerance at scale.}



\section{Background}
\label{sec:background}

\subsection{Agentic RL Training}
\label{subsec:multi-task-agentic}

\PHB{The Training Pipeline.}
Multi-task agentic RL training follows an iterative loop with three stages.
The first stage, \textbf{rollout}, collects experience: an agent LLM (actor)
interacts with parallel \emph{environments} to generate training data. Unlike
standard LLM inference, rollout is \emph{multi-turn} and \emph{stateful}~\cite
{kimiteam2025kimik2openagentic,deepseek_v3_2_speciale}. At each turn, the
agent observes a state, emits an action token sequence, and submits it to the
environment. The environment executes the action (e.g., running code or
clicking a link) and returns feedback. The loop repeats until termination,
producing a sequence of state-action pairs, i.e., a \emph{trajectory}.
After rollout, the \textbf{reward stage} evaluates trajectory quality via a
\emph{reward worker}, producing a scalar reward signal. Reward computation
ranges from lightweight rule-based checks~\cite{he2025deepmath} to
computationally expensive model-based judgments (e.g., LLM-as-a-Judge~\cite
{son2024llmasajudgerewardmodel,zhong2024rlhfuseefficientrlhftraining}).
Finally, the \textbf{training stage} consumes trajectories and rewards to
update model weights with RL algorithms (e.g., PPO~\cite{schulman2017proximal},
GRPO~\cite{shao2024deepseekmath}). To attain optimal training performance~\cite{areal},
production RL pipelines often adopt \textit{synchronous} training, which enforces
strict weight synchronization between rollout and training at every step.

\begin{table}[t]
\centering
\caption{Taxonomy of Adopted Agentic Environments.}
\label{tab:env-taxomy}
\resizebox{0.45\textwidth}{!}{%
\begin{tabular}{ll l c}
\toprule
\textbf{Environment} & \textbf{Task} \textbf{Domain}   & \textbf{Modality}     & \textbf{\#Turns} \\
\midrule
SWE-bench~\cite{swe-bench}   & SWE           & Text         & 30--50      \\
WebShop~\cite{webshop}     & Web           & Text         & 5--30      \\
FrozenLake~\cite{frozen_lake}  & Game          & Text, Visual & 20--100      \\
GEM-math~\cite{GEM-github}    & Math+Tool Use & Text       & $<5$         \\
GEM-game~\cite{GEM-github}    & Game          & Text         & 1            \\
\bottomrule
\end{tabular}}
\end{table}

\begin{figure}[tb]
    \centering      
    \begin{subfigure}{0.48\textwidth}
        \centering
        \includegraphics[width=\linewidth]{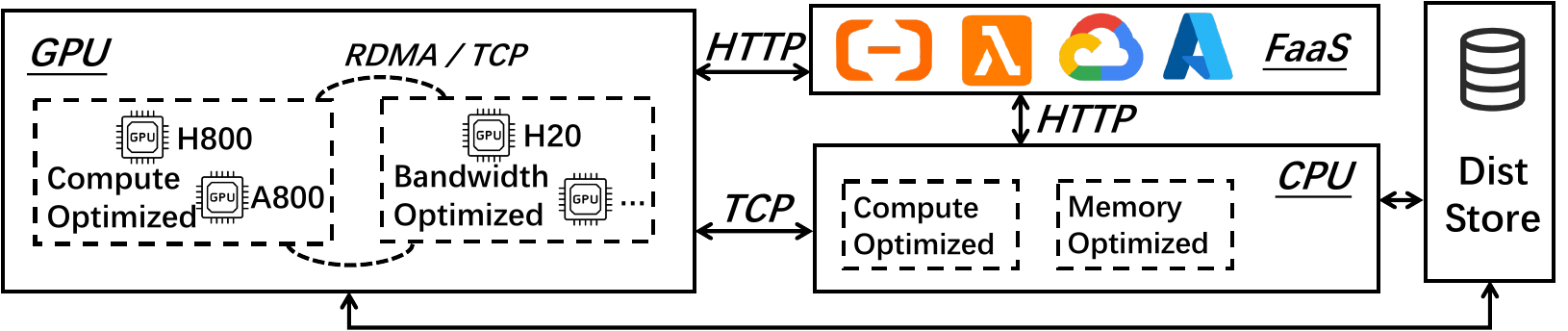}
    \end{subfigure}    
    \caption{
        Disaggregated infrastructure for agentic RL.
    }
    \label{fig:disaggregated_cluster}
\end{figure}

\PHM{Environment Heterogeneity.}
A key challenge in agentic RL is environment diversity, which determines the
system compute profile. As summarized in \autoref{tab:env-taxomy}, agentic tasks vary
widely in modality and interaction frequency. Complex reasoning tasks such as
\texttt{SWE-bench}~\cite
{swe-bench} (software engineering), \texttt{FrozenLake}~\cite{frozen_lake}
(visual games), and \texttt{WebShop}~\cite{webshop} (eCommerce) require long
interaction horizons (up to 30--100 turns). Frequent interactions force the
agent to repeatedly process growing context histories, making these workloads
prefill-heavy and compute intensive. In contrast, tasks like \texttt{GEM-Math}
and \texttt{GEM-Game}~\cite{GEM-github} may involve fewer turns (up to five)
but require longer chains of thought per action. These workloads are
decoding-heavy, shifting the bottleneck from compute to memory bandwidth. This
variance requires infrastructure that adapts to each task's rollout profile.

\begin{figure}[tb]
    \centering      
    \begin{subfigure}{0.48\textwidth}
        \centering
        \includegraphics[width=\linewidth]{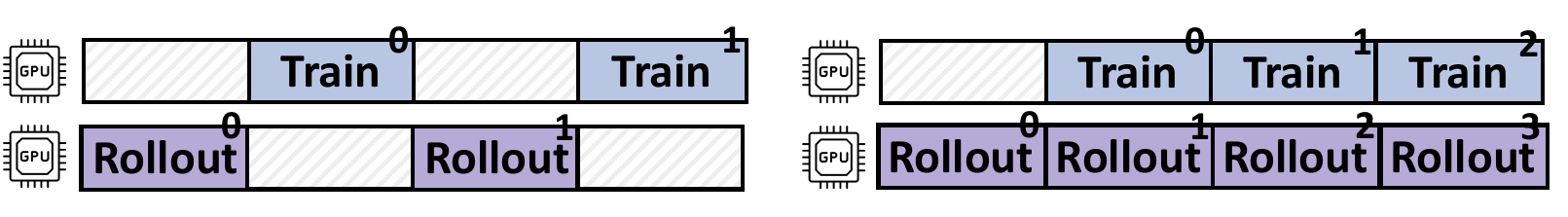}
    \end{subfigure}    
    \vspace{-10pt}
    \caption{
        Synchronous vs. asynchronous training.
    }
    \label{fig:paradigm-all-in-one}
\end{figure}

\subsection{Disaggregated Cluster Infrastructure}
\label{subsec:disaggregated-cluster}

\PHB{The Case for Disaggregation.} The extreme heterogeneity of agentic RL
workloads, ranging from compute-intensive training to stateful environment
simulation (\S\ref{subsec:multi-task-agentic}), renders monolithic architectures
inefficient. Consequently, agentic RL training must transition to a \emph
{disaggregated infrastructure} that decouples these demand-conflicting
computation stages into specialized resource pools. As shown in
\autoref{fig:disaggregated_cluster}, in a typical disaggregated infrastructure,
\emph{training clusters} utilize high-end, compute-optimized GPUs (e.g., NVIDIA
H800) to sustain high throughput; \emph{inference clusters} leverage
bandwidth-optimized hardware (e.g., H20) to serve memory-bound decoding;
\emph{CPU clusters} provide elastic capacity for diverse, containerized runtime
environments orchestrated by Kubernetes~\cite{k8s}; and \emph{serverless
infrastructure} handles bursty, stateless workloads like reward evaluation.
These pools are interconnected through standard network fabrics, relying on
distributed storage for persistent logging and fault tolerance.

\PHM{Sync. vs. Async. Training.} 
While disaggregation resolves resource mismatches, it introduces non-trivial
orchestration challenges to the training paradigm that dictates 
the trade-off between system throughput
and algorithmic consistency.

\underline{\emph{1) Synchronous Training:}}
This paradigm enforces strict consistency by blocking rollout until it receives
the latest model weights from the training cluster. In disaggregated settings,
this causes substantial ``dependency bubbles'' (\autoref
{fig:paradigm-all-in-one}-Left), where expensive GPUs sit \emph{idle} during
high-latency weight synchronization and straggler-bound environment steps
(\S\ref{sec:motivation}).

\underline{\emph{2) Asynchronous Training:}}
To mitigate these bubbles, systems can adopt \emph{asynchronous} paradigms
(e.g., one-off RL training~\cite{deepscaler2025} in \autoref
{fig:paradigm-all-in-one}-Right). Rollout and training then run \emph{in
parallel}: training consumes trajectories from slightly older policies (e.g.,
one iteration stale in one-off training), while rollout continuously produces
new data. This design masks synchronization latency and straggler effects
inherent to disaggregation, trading some policy staleness for higher hardware
utilization and throughput.

\begin{table}[t]
\centering
\caption{NVIDIA GPU specifications.}
\label{tab:nvidia-gpu-specs}
\small
\begin{tabular}{lcc}
\toprule
Hardware Specification                & H20            & H800           \\
\midrule
TFLOPS         & 148            & 989.5 \\
HBM capacity        & 96GB  & 80GB           \\
HBM bandwidth       & 4TB/s & 3.35TB/s       \\
NVLink bandwidth    & 900GB/s & 400GB/s      \\
Normalized Cost~\cite{megascale-infer} & 1.00          & 2.85 \\
\bottomrule
\end{tabular}
\end{table}

\section{Characterization and Requirements}
\label{sec:motivation}

To motivate the design of \SystemName{}, we conduct a comprehensive workload
characterization of multi-task agentic RL training. Based on these empirical
observations, we derive four critical system requirements that \SystemName{}
must satisfy: within-generation hardware affinity mapping (\textbf{R1}),
trajectory-level asynchrony (\textbf{R2}), serverless reward
offloading (\textbf{R3}), and bounded-staleness asynchronous training
(\textbf{R4}).

\subsection{Stage Computation}
\label{subsec:resource-req-within-stage}

\PHB{Training Step Latency Breakdown.} 
We first profile the end-to-end latency of a standard training iteration to
identify the dominant cost components. We train Qwen3-8B/32K on 32 H800 GPUs
using the \texttt{SWE-bench} environment (batch size 128), where the agent LLM
interacts with a containerized sandbox for software engineering tasks. The
interaction involves two core operations: \texttt{env.reset} for environment
initialization via Docker image pulling and container launching, and \texttt{env.step}
for agent's action execution.

\begin{figure}[t]
    \centering      
    \includegraphics[width=\linewidth]{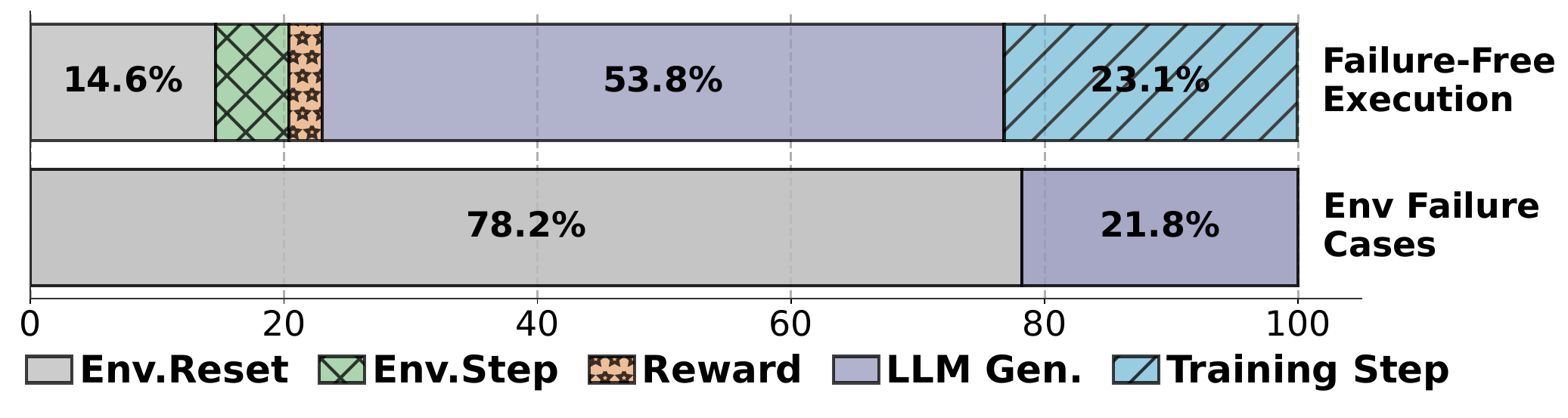}  
    \caption{
       Breakdown of a training step: successful runs (top, avg=365.7s) versus execution with environment failures (bottom, avg=513.3s). 
    }
    \label{fig:time_breakdown}
\end{figure}

\autoref{fig:time_breakdown} breaks down the latency of five \emph
 {successful} iterations against five iterations containing \emph
 {environment failures}. Even on the success path, the average iteration time is 366 seconds, with LLM generation accounting for only 54\%. The rest is split across training (23\%), environment initialization (15\%), and other overheads, a mixed profile that invalidates the common assumption of overwhelming generation dominance~\cite{rollpacker,RhymeRL}. When
 environment timeouts occur, the average time spikes to 513 seconds and
 \texttt{env.reset} alone consumes 78\% of rollout time, shifting the
 bottleneck entirely from GPU computation to environment overhead. Our
 production data indicates these failures are not rare corner cases, occurring
 approximately once every ten iterations (\S\ref{sec:large-scale-analysis}). Our empirical study therefore reveals that, in agentic RL, the system must handle four heterogeneous workloads, including generation, training, environment, and reward, rather than optimizing for the generation dominant regime reported by prior systems.


\begin{figure}[tb]
    \centering
    \begin{subfigure}{0.235\textwidth}
        \centering
        \includegraphics[width=\linewidth]{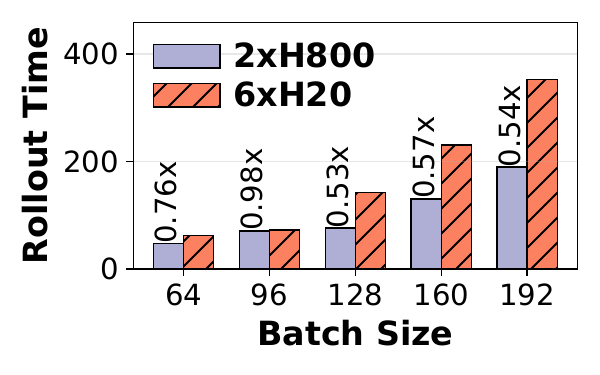}
        \caption{\texttt{FrozenLake} \small{[Prefill-Heavy]}.}
        \label{fig:prefill-heavy-task}
    \end{subfigure}
    \begin{subfigure}{0.235\textwidth}
        \centering
        \includegraphics[width=\linewidth]{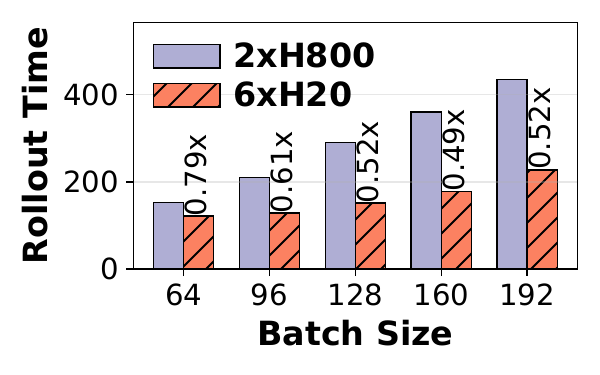}
        \caption{\texttt{GEM-Math} \small{[Decode-Heavy]}.}
        \label{fig:decode-heavy-task}
    \end{subfigure}
    \caption{End-to-end rollout time (seconds) of different tasks on H20 and H800 GPUs across varying batch sizes.}
    \label{fig:motivation-task-affinity}
\end{figure}

\PHM{Divergent Hardware Affinities in Generation.} 
Modern GPUs expose different trade-offs between compute capability, memory
capacity, and cost (\autoref{tab:nvidia-gpu-specs}). While recent RL
systems~\cite{StreamRL,asyncflow,seamlessflow} advocate physically decoupling
generation from training, assigning rollout to cost-effective,
bandwidth-optimized GPUs and training to
compute-optimized GPUs, our characterization reveals that \emph{this
static assignment is insufficient for agentic workloads}. LLM generation
comprises two distinct phases with opposing resource demands: the 
compute-bound prefill phase and the memory-bandwidth-bound decoding phase. In
environments with a few interaction turns but long chains of thought per
action, most of the runtime is spent in decoding (decoding-heavy); conversely,
in environments with many turns, the prefill phase dominates and demands high
compute throughput (prefill-heavy). In our production clusters,
agentic RL tasks exhibit a clear \emph{bimodal distribution}, featuring 
either a small number of interaction turns ($< 5$) or a large number ($> 10$).

To quantify this divergence, we run a prefill-heavy task (\texttt
{FrozenLake}) and a decoding-heavy task (\texttt{GEM-Math}) using
Qwen3-8B/32K for ten iterations with prefix caching enabled. For a
cost-equivalent comparison, we execute the workloads on two distinct hardware
configurations: one with six H20 GPUs and the other with two H800 GPUs. As shown
in \autoref{fig:prefill-heavy-task}, the compute-heavy H800 outperforms the
H20 on \texttt{FrozenLake}, reducing end-to-end rollout time to as low as 0.53$\times$. Conversely, for \texttt{GEM-Math} (\autoref{fig:decode-heavy-task}), the
H20's higher memory bandwidth accelerates decoding, reducing rollout time to
0.49$\times$--0.79$\times$ of the H800. These
results invalidate the assumption that generation is uniformly
bandwidth-bound. Instead, maximizing throughput requires 
dynamically mapping tasks to their best-fit hardware.

\begin{tcolorbox}[colback=blue!5!white,colframe=gray!75!black,left=1mm, right=1mm, top=0.5mm, bottom=0.5mm, arc=1mm]
\noindent\textbf{R1:} Generation is not uniformly bandwidth-bound; the system must bind each task
to a best-fit GPU class rather than committing rollout to a single GPU type.
\end{tcolorbox}

\begin{figure}[tb]
    \centering      
    \begin{subfigure}{0.2\textwidth}
        \centering
        \includegraphics[width=\linewidth]{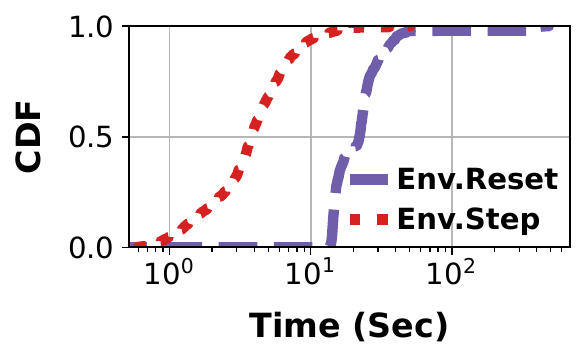}
        \caption{Env Time Distribution.}
        \label{fig:env_time_cdf}
    \end{subfigure}
    \begin{subfigure}{0.27\textwidth}
        \centering
        \includegraphics[width=\linewidth]{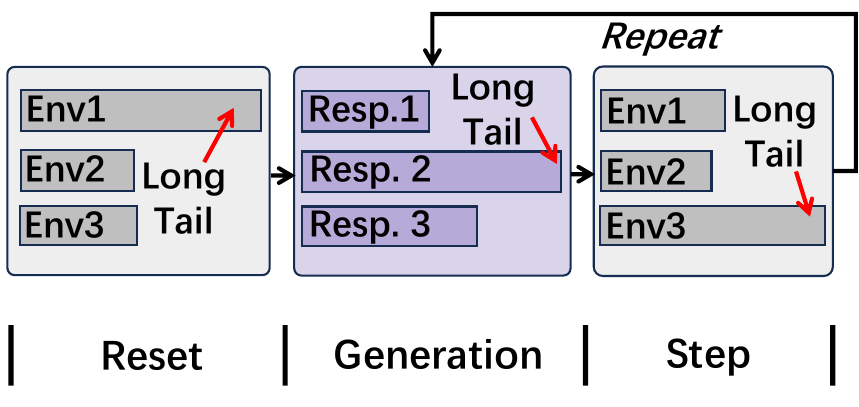}
        \caption{Batched Env Interaction.}
        \label{fig:sub-batched_env_interaction}
    \end{subfigure}
    
    \caption{The analysis of environment interaction: (a) Cumulative distribution function of time (log-scaled) taken for environment initialization (\texttt{env.reset}) and environment step (\texttt{env.step}). (b) Illustration of how long-tail environments affect multi-turn rollouts under batched env interaction.
    }
    \label{fig:env_analysis}
    \vspace{-5pt}
\end{figure}

\PHB{Heavy-Tailed Environment Execution.} 
\yuheng{
Since each training step collects a batch of trajectories, it runs hundreds to
thousands of environments \emph{simultaneously}, at which scale strong
resource isolation becomes essential: without it, concurrent disk I/O
exhausts shared quotas and triggers cascading failures. We therefore employ
Kubernetes clusters to manage and isolate each containerized environment.
}


\yuheng{
A substantial body of prior
work~\cite{rollpacker,StreamRL,RhymeRL} has
analyzed the long-tail behavior of LLM generation. 
In agentic RL, environment interactions introduce
additional sources of long-tail execution patterns.
\autoref{fig:env_time_cdf} illustrates the latency
distributions of \texttt{env.reset} and
\texttt{env.step}, both exhibiting pronounced
long tails.
The long-tail delay of \texttt{env.reset} can reach
hundreds of seconds in production, mainly due to
(1)~network contention, where 
concurrent Docker image pulls
saturate network links, and
(2)~compute and I/O contention on host nodes, where
launching containers consumes substantial CPU and
disk resources.
The cumulative \texttt{env.step} time per trajectory
also varies widely, driven by the large variance in
interaction turns and per-step overhead.}


These long-tail trajectories act as stragglers that delay end-to-end rollout
latency. Since LLM engines execute requests in batches, it is natural to batch
environment interactions with the agent LLM as well. 
\autoref{fig:sub-batched_env_interaction} illustrates this pattern: fast
environments must wait for the slowest one before the next generation step can
proceed. Our profiling (\autoref{fig:time_breakdown}) indicates that batched
environment interaction increases rollout time by up to 21.3\% compared to ideal
execution, an overhead that compounds as environment failure rates increase.
\yuheng{Trace analysis on our production deployment
(\S\ref{sec:large-scale-analysis}) confirms that such long-tail
conditions are not rare throughout training, and would
significantly penalize any batched execution scheme.}


\begin{tcolorbox}[colback=blue!5!white,colframe=gray!75!black,left=1mm, right=1mm, top=0.5mm, bottom=0.5mm, arc=1mm]
\textbf{R2:} Environment execution, including \texttt{env.reset} and
\texttt{env.step}, is prone to extreme long-tail latency. The system must manage
environment execution at trajectory granularity rather than batch granularity,
so that slow or failed environments do not stall the pipeline.
\end{tcolorbox}

\begin{table}[t]
\centering
\caption{\small Transmission overhead from the training cluster to the inference cluster over TCP and RDMA.}
\label{tab:llm-transmission}
\scriptsize
\resizebox{0.45\textwidth}{!}{%
\begin{tabular}{lcccc}
\toprule
\textbf{Model} & \textbf{Size (GB)} & \textbf{TCP (s)} & \textbf{RDMA (s)} & \textbf{Speedup} \\
\midrule
Qwen3-8B & 15.26  & 6.911  & 5.466  & 1.264$\times$ \\
Qwen3-14B & 27.51  & 14.437  & 5.817  & 2.482$\times$ \\
Qwen3-32B & 61.02  & 29.649 & 9.442 & 3.140$\times$ \\
\bottomrule
\end{tabular}
} 
\end{table}

\begin{figure}[tb]
    \centering      
    \includegraphics[width=\linewidth]{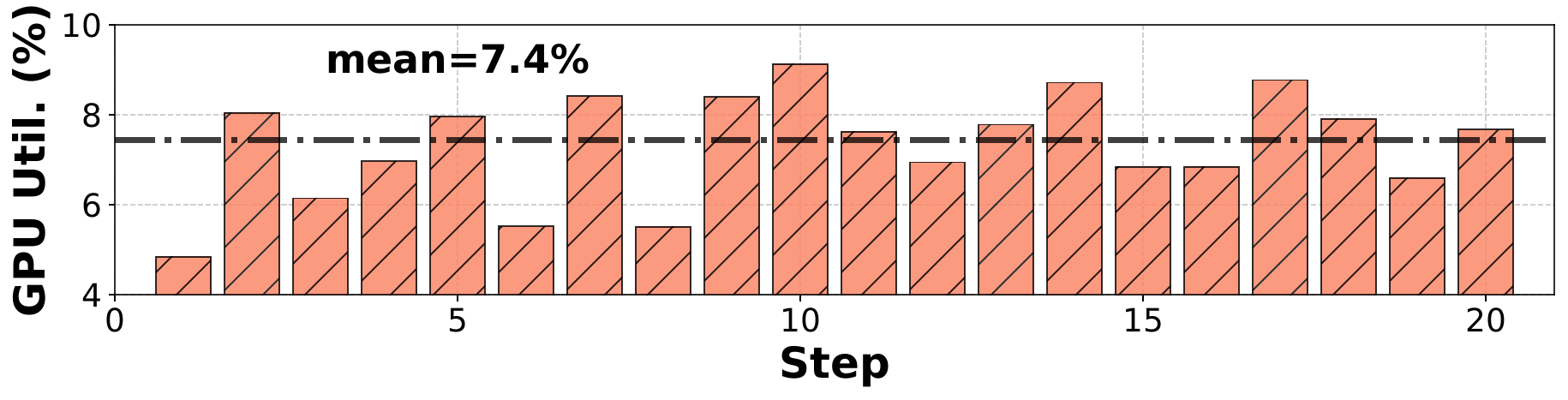}  
    \vspace{-15pt}
    \caption{
        Inefficient resource usage when dedicating local GPUs for reward computation.
    }
    \vspace{-5pt}
    \label{fig:reward_utilization}
\end{figure}

\PHB{Stateless Reward Computation.} The reward stage follows rollout, and most
reward computations can be implemented as \emph{stateless functions}. For
lightweight \emph{rule- and code-based rewards}, they have small, bursty
resource demand, making them a natural fit for elastic serverless execution that
can scale to zero between steps. The \emph{LLM-based reward computation}, on the
other hand, demands intensive GPUs: colocating them with rollout contends with
generation for GPU memory, KV-cache capacity, and scheduling slots, degrading
both stages under concurrent demand. A common workaround is to
reserve separate GPUs for the reward LLM: in our Qwen3-8B/32K
\texttt{SWE-bench} run with batch size 128, we allocate four H800 GPUs to a
dedicated 7B reward LLM and 28 H800 GPUs to rollouts, but the reward GPUs
achieve only 7.4\% average utilization across steps
(\autoref{fig:reward_utilization}).
Since the reward LLM's parameters remain fixed during training, it can be
treated as a stateless function, and serverless
deployment~\cite{Torpor,ServerlessLLM,BLITZSCALE} simultaneously avoids
rollout contention and reclaims the idle GPU budget wasted by dedicated
reservations.

\begin{tcolorbox}[colback=blue!5!white,colframe=gray!75!black,left=1mm, right=1mm, top=0.5mm, bottom=0.5mm, arc=1mm]
\noindent\textbf{R3:} Reward workers are stateless, bursty, and persistently
underutilized; colocating them with rollout
contends for GPU memory under concurrent generation, making elastic serverless
deployment the better fit.
\end{tcolorbox}

\subsection{Inter-Stage Communication}
\label{subsec:communication-between-stage}

Beyond computation, inter-stage communication plays a critical role in agentic
RL training, comprising two distinct types: stability-critical, small-packet
\emph{trajectory transfer} and bandwidth-intensive, large-volume \emph{weight
update}.

\PHM{Stability-Critical Trajectory Transfer.} Trajectory and observation
payloads are orders of magnitude smaller than the multi-GB weight transfers: a
single agent-environment exchange typically carries kilobytes to
a few megabytes of tokenized state and action data, yet each rollout issues
hundreds to thousands of such exchanges per trajectory. Consequently, cumulative
latency, rather than bandwidth, determines end-to-end interaction cost. In practice,
environment interaction should therefore prioritize \emph{network stability}
over \emph{network bandwidth}, and asynchronous execution between environment
interaction and LLM generation prevents network latency from becoming the
rollout bottleneck.

\PHM{Bandwidth-Intensive Weight Update.} During training, the agent LLM
periodically updates its weights, which must then be synchronized with the
rollout stage. This weight synchronization is the dominant source of inter-stage
communication overhead. We measure the end-to-end transmission cost of
synchronizing model parameters between the training and inference clusters using
Mooncake~\cite{qin2024mooncake} over TCP (200 Gbps Ethernet) and RDMA (400 Gbps
InfiniBand), and report the results in \autoref{tab:llm-transmission}. RDMA
provides higher bandwidth and lower communication overhead than TCP. In
\emph{synchronous} RL training, the rollout stage can only proceed after the
latest agent LLM weights have been synchronized. As a result, the substantial
cost of weight transmission over low-bandwidth links increases end-to-end
training time and diminishes the speedup of disaggregated training
(\autoref{fig:paradigm-all-in-one}-Left).

In asynchronous training (\autoref{fig:paradigm-all-in-one}-Right),  the
training and rollout stages execute in parallel on separate GPUs. Although this
introduces data staleness, many prior works~\cite{rollflash,RhymeRL}
empirically observe that asynchronous training can preserve model quality
under a staleness bound. Given the dominant rollout overhead, asynchronous
training can effectively hide both training and weight synchronization costs
with rollout, thereby reducing end-to-end training latency.

\begin{tcolorbox}[colback=blue!5!white,colframe=gray!75!black,left=1mm, right=1mm, top=0.5mm, bottom=0.5mm, arc=1mm]
\textbf{R4:} Training and rollout must execute concurrently on separate GPU
clusters with \emph{bounded} staleness for system efficiency and training quality.
\end{tcolorbox}

\section{System Overview}
\label{sec:overview}

We present \SystemName{}, a distributed system that orchestrates the rollout,
reward, and training stages of agentic RL across disaggregated, heterogeneous
resource pools. \SystemName{} is implemented in approximately 60k lines of
Python code. We first present the system architecture and then walk through one
training iteration to illustrate its workflow.

\subsection{Architecture}
\label{subsec:arch-overview}

\begin{figure}[t]
    \centering
    \includegraphics[width=0.8\linewidth]{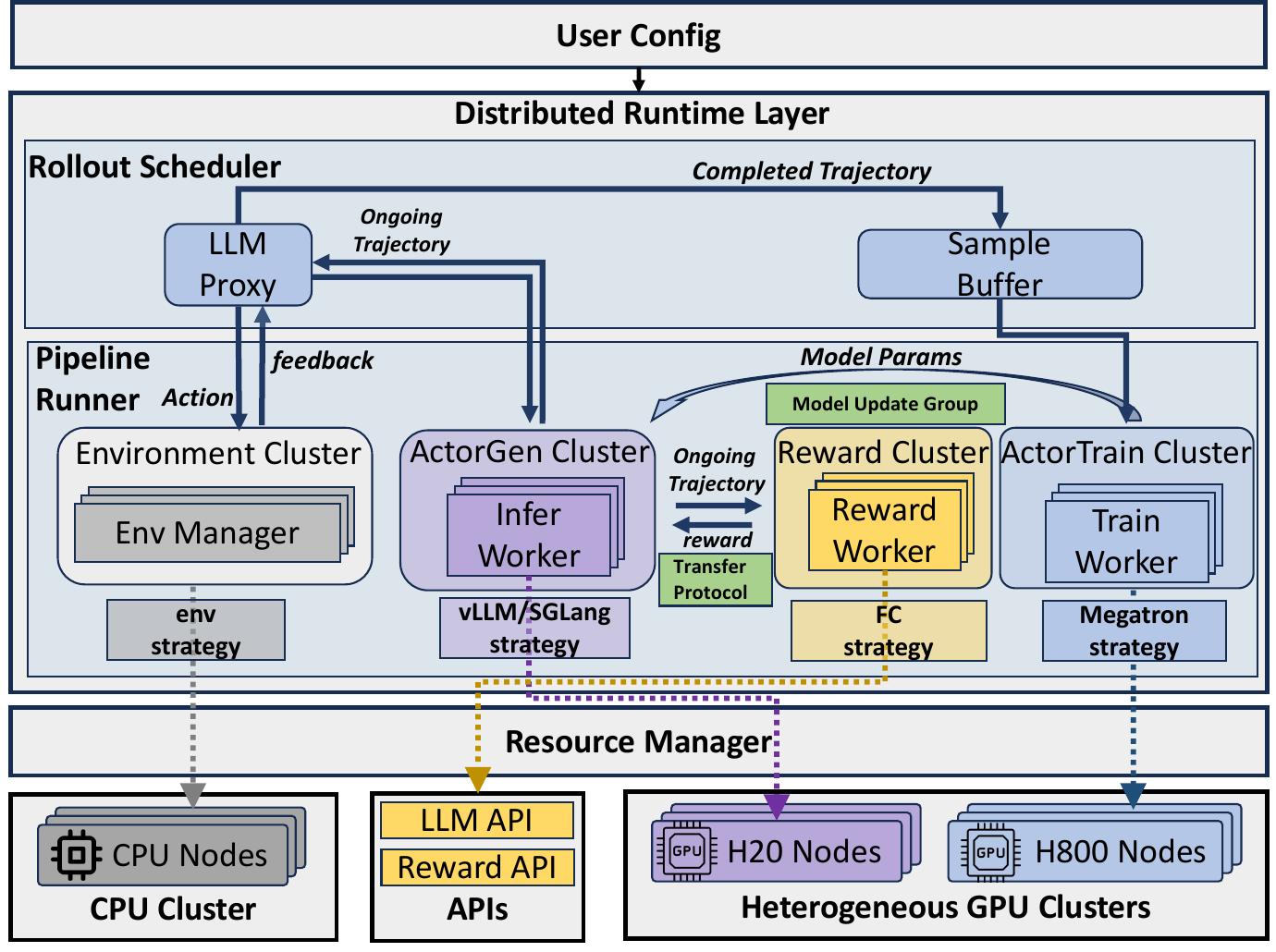}
    \caption{System Architecture of \SystemName{}.}
    \label{fig:sys_arch}
    \vspace{-15pt}
\end{figure}

To meet the requirements in \S\ref{sec:motivation}, \SystemName{} is
organized into three layered planes: resource, data, and control, as shown in
\autoref{fig:sys_arch}. The \emph{resource plane} decides placement: it tracks
heterogeneous hardware pools and binds each role, task, or generation sub-phase
to a compatible resource class using hardware-affinity declarations
(\textbf{R1}). The \emph{data plane} realizes the pipeline through
\texttt{Worker} and \texttt{Cluster} abstractions that encapsulate user-defined
execution logic and offload stateless computation to
serverless infrastructure (\textbf{R3}). The \emph{control
plane} coordinates progress: it schedules trajectories independently, dispatches
reward computation as trajectories finish, buffers scored samples, and overlaps
training with rollout under a bounded-staleness protocol (\textbf{R2} \&
\textbf{R4}).

\PHM{Resource Plane.} The resource plane is implemented by the \emph{resource
manager} (\autoref{fig:sys_arch}), which monitors the state of disaggregated
hardware pools and binds resources to specific \texttt{Worker}s based on their
roles and corresponding hardware affinities (e.g., training to H800 GPUs,
rollout to H20 GPUs, environment execution to CPU servers, and reward
computation to serverless infrastructure).

\PHM{Data Plane.} The data plane executes the distributed RL job on resources
provisioned by the resource plane. It is implemented by the \emph{pipeline
runner}, which constructs role-specific \texttt{Cluster}s for training,
generation, environment execution, and reward computation. Each \texttt{Cluster}
then launches its associated \texttt{Worker}s and orchestrates their execution
over the appropriate backend runtime (e.g., Megatron~\cite{shoeybi2019megatron}
and vLLM~\cite{vllm}). In this way, the data plane translates user-defined
\texttt{Worker}/\texttt{Cluster} logic into concrete distributed execution.

\PHM{Control Plane.} The control plane coordinates trajectory-level rollout and
asynchronous training. It is centered on the \emph{rollout scheduler}, which
orchestrates the interaction among environment execution, generation workers,
reward workers, and training workers. To decouple these stages, \SystemName{}
introduces two auxiliary components: \texttt{LLMProxy}, which mediates
generation requests between environment execution and generation \texttt{Worker}s, and
\texttt{SampleBuffer}, which buffers completed and scored trajectories for
training. Together, these components allow rollout, reward evaluation, and model
updates to proceed concurrently, hiding trajectory-level synchronization and
weight-update coordination from the user.

\subsection{End-to-End Workflow}
\label{subsec:e2e-workflow}
To illustrate how the three planes interact, we walk through the first asynchronous
training iteration in \SystemName{}. At startup, the resource manager in the
resource plane binds heterogeneous GPU, CPU, and serverless resources to the
corresponding \texttt{Cluster}s by allocating and placing their constituent
\texttt{Worker}s according to user-specified hardware affinities. The pipeline runner
then materializes the data plane by instantiating the corresponding
\texttt{Cluster}s for training, generation, environment execution, and reward
computation, each backed by its assigned \texttt{Worker}s and execution engine.

Once the pipeline is instantiated, the rollout scheduler in the control plane
launches multiple \texttt{EnvManager} instances to collect trajectories. Each
\texttt{EnvManager} alternates between issuing generation requests through the
shared \texttt{LLMProxy} and applying the returned actions to its environment.
Completed trajectories are scored by a reward backend and deposited into the
\texttt{SampleBuffer}, from which the training \texttt{Cluster} consumes batches
and updates the model. Rollout, reward evaluation, and training proceed
concurrently. 

The following sections detail these planes. \S\ref{sec:programming-model}
describes the programming abstractions and mechanisms that configure the
resource and data planes. \S\ref{sec:runtime} describes the system-managed
control plane that drives runtime execution, followed by cross-cutting
optimizations that span multiple planes.

\lstset{basicstyle=\scriptsize\ttfamily}
\section{Programmable Resource and Data Planes}
\label{sec:programming-model}

\SystemName{} provides a declarative programming model that separates user
intent from system execution. Users specify execution logic and
task-specific mappings for the data plane, as well as hardware preferences for
the resource plane, through the \texttt{Worker}-level interface. The resource
manager interprets the resource-plane declarations to construct concrete
bindings over heterogeneous resources, while the pipeline runner materializes
the data plane through \texttt{Cluster} abstractions.

\subsection{\texttt{Worker} and \texttt{Cluster} Abstractions}
\label{sec:abstractions}

The resource and data planes are built around two core abstractions:
\texttt{Worker} and \texttt{Cluster}.

\PHM{\texttt{Worker} Abstraction.} A \texttt{Worker} is the basic execution
unit that spans both the resource and data planes. It encapsulates user-defined
computation logic together with method-level declarations, and executes on the
hardware provisioned by the resource plane. A \texttt{Worker} can be specialized into the roles required across stages, including training, generation, reward, and environment, by subclassing and method annotations.

\PHM{\texttt{Cluster} Abstraction.} While \texttt{Worker}s execute the
actual computation logic, managing thousands of \texttt{Worker}s individually is
prohibitive for algorithm developers. To address this, \SystemName{} introduces the
\texttt{Cluster} abstraction. A \texttt{Cluster} acts as a proxy and controller
for a role-specific \texttt{Worker} group. A \texttt{Cluster} spawns
\texttt{Worker}s, binds their methods, and manages collective invocation on
their behalf. In practice, \SystemName{} defines four \texttt{Cluster}s: \texttt{ActorTrain}, \texttt{ActorGen}, \texttt{Reward}, and \texttt{Environment}, corresponding to the four RL stages shown in \autoref{fig:sys_arch}. By composing these abstractions, \SystemName{} maps the agentic RL pipeline onto the disaggregated hardware fabric.

\begin{figure}[t]
\centering
\begin{lstlisting}[
    caption={Worker-level programming interface of \SystemName{}.},
    label={list:api-example-1}
]
import rollart.distributed as rdist
from rdist.worker import ActorTrainCls, ActorGenCls, RewardCls
from rdist import ResourceManager as RM

# 1. Single Controller Example
class MyActorTrain(ActorTrainCls):
    @rdist.register(mode="execute_all")
    def compute_gradients(self, input_tensor):
        ...

# 2. Define actor_gen on heterogeneous GPUs
# 2.1 heterogeneous GPU allocation. 
gen_rm=RM( {"H800": list(range(0, 8))},  
           {"H20", list(range(8, 32))})
# 2.2 hardware affinity mapping
class HeteroActorGen(ActorGenCls):
    @rdist.hw_mapping(
    hw_affinity={"FrozenLake": "H800","default": "H20"}
    )
    def generate(self, input_ids:List[int], 
                tag_name:str="default"):
        return self.model.process(prompt)

# 3. Define a serverless reward computation func
class ServerlessRewardWorker(RewardCls):
    @rdist.register_serverless(
        attribute='reward_proxy',
        serverless_url='fc://xxx.xxx')
    def compute_rewards(self, traj: list):
        prompt = f"Evaluate the trajectory:{traj}"
        return ray.get(self.reward_proxy(prompt))

    
\end{lstlisting}
\vspace{-10pt}
\end{figure}

\subsection{\texttt{Worker} Declarations and Binding}
\label{sec:worker-interface}

\SystemName{} exposes three worker-level interfaces through Python decorators,
allowing users to implement role-specific computation logic and specify
task-specific mappings and hardware preferences under a unified
\emph{single-controller} model. The resource manager interprets the resource-related
declarations to bind \texttt{Worker}s to heterogeneous resources.

\PHM{Decorator-Based Interface.} \autoref{list:api-example-1} shows the three
decorator-based interfaces for \texttt{Worker} declarations.

\emph{\uline{1) Single Controller.}} Like other industry RL
frameworks~\cite{verl,slime_github}, \SystemName{} adopts the single controller
programming model to streamline pipeline construction. When a \texttt{Worker}
method is annotated with the \texttt{register} decorator and set to
\texttt{execute\_all} mode (Lines~7--8 in \autoref{list:api-example-1}), the
runtime broadcasts inputs and invokes the method across all \texttt{Worker}s in
the corresponding \texttt{Cluster}. The runtime also collects the returned
results and automatically aggregates them. A typical use case is defining how
gradients are computed on each training \texttt{Worker} (Line~8 in
\autoref{list:api-example-1}).

\emph{\uline{2) Hardware-Affinity Mapping.}} To align workload characteristics
with hardware capabilities (\textbf{R1}), \SystemName{} allows users to declare
preferred hardware targets for role-specific workers through the
\texttt{hw\_mapping} decorator. By default, training workers (\texttt{ActorTrainCls}) are mapped to
compute-optimized GPUs (e.g., H800), generation workers (\texttt{ActorGenCls})
to bandwidth-optimized GPUs (e.g., H20), environment workers
(\texttt{EnvironmentCls}) to CPU servers, and reward workers
(\texttt{RewardCls}) to local GPU servers. Users can override these defaults
with finer-grained preferences.

As shown in the \texttt{HeteroActorGen} class, a dictionary-based resource
specification provisions heterogeneous GPU groups (Lines~13--14), and the
\texttt{hw\_mapping} decorator (Lines~17--19) declares sub-stage affinities.
Prefill-heavy \texttt{FrozenLake} rollouts are routed to H800 GPUs, while all
other tasks default to H20s. At runtime, the rollout scheduler passes the
current task's identity (e.g., \texttt{FrozenLake}) as the \texttt{tag\_name}
argument (Line~21), allowing the system to route each generation request to the
most appropriate hardware. 


This declaration is coarse-grained by design: users specify affinity
at the task-domain level rather than performing per-request load-balancing. In
our production traces, different agentic task domains diverge in computation characteristics (e.g., turn counts and prefill/decode ratios),
making domain labels a practical basis for hardware selection
(\S\ref{sec:large-scale-analysis}). Our evaluation confirms this
lightweight annotation captures enough heterogeneity to improve throughput
(\S\ref{subsec:eval-ablations}). 
This leaves affinity selection explicit, and \S\ref{sec:discussion} discusses how an online profiler could automate or refine these mappings at runtime.

\emph{\uline{3) Serverless Registration.}} Reward workers are stateless and
exhibit low utilization on dedicated GPUs (\textbf{R3}).
To address this, \SystemName{} provides the \texttt{register\_serverless}
decorator (Lines~26--28) to offload reward computation to serverless
platforms. With the specified
\texttt{serverless\_url}, the runtime invokes \texttt{compute\_rewards}
(Line~29) as a pure function through the same method-level interface while
redirecting execution to an external serverless platform (e.g., Function Compute~\cite{alibaba_function_compute}). This grants the system
zero-overhead autoscaling without provisioning costly, hot-standby
GPUs.

\PHM{Resource Binding.} The resource manager uses a shared metadata
store (e.g., Redis) to maintain a global, real-time view of resource pools,
including compute-optimized GPUs (e.g., H800), bandwidth-optimized GPUs (e.g.,
H20), CPU clusters, and serverless endpoints. \texttt{Worker}-level annotations specify
hardware affinities and execution targets for individual methods. Upon a \texttt{Worker}
deployment request, the resource manager interprets these declarations to
determine concrete placements and bindings. It first checks the availability of
the preferred resource pool and binds the requested \texttt{Worker} accordingly. If the
preferred hardware is temporarily unavailable, the manager opportunistically
falls back to compatible default resources rather than stalling deployment. The
resulting binding metadata is recorded for subsequent dispatch, failover, and
reconfiguration.


\begin{figure}[t]
\centering
\begin{lstlisting}[
    caption={Simplified implementation of \texttt{Cluster}.},
    label={list:api-example-cluster}
]
class Cluster: 
    def __init__(self, res_manager, worker_cls): 
        self._create_worker(worker_cls, res_manager)
        self._bind_worker_method()
    
    def execute_all(self, method_name, *args, **kwargs):
        result = []
        for worker in self.workers: 
            rcall = getattr(worker, method_name)
            result.append(rcall(*args, **kwargs))
        return ray.get(result)

    def hw_mapping(self, hw_affinity, tag_name, *args):
        hw_type = hw_affinity.get(tag_name)
        new_workers = []
        for worker in self.workers:
            if worker.resource_type == hw_type: 
                new_workers.append(worker)
        # route requests to new_workers next

    def register_serverless(self, attr, url, *args): 
        # define a call_fc to call serverless url
        for worker in self.workers: 
            setattr(worker, attr, call_fc)
        # perform execute_all logic next
\end{lstlisting}
\vspace{-5pt}
\end{figure}

\subsection{\texttt{Cluster}-Level Realization}
\label{sec:cluster}

The resource manager realizes the resource plane, while the \texttt{Cluster}
governs execution in the data plane. We next describe how the default
\texttt{Cluster} implementation turns \texttt{Worker}-level decorators into concrete
execution behavior.

\PHM{\texttt{Cluster} Construction and Method Binding.} At initialization,
\texttt{Cluster} assembles a role-specific \texttt{Worker} group from the
specified \texttt{worker\_cls} using resources provisioned by the resource
manager, and initializes the corresponding backend runtime (e.g., Megatron or
vLLM) so that \texttt{Worker} methods execute on that backend (Line~3 in
\autoref{list:api-example-cluster}). Each \texttt{Worker} is associated with
resource metadata, such as its hardware type, enabling later affinity-aware
dispatch. The \texttt{\_bind\_worker\_method} function then binds each method of
\texttt{worker\_cls} to the \texttt{Cluster} instance, allowing the
\texttt{Cluster} to act as an invocation proxy (Line~4). For example, if
\texttt{worker\_cls} defines \texttt{compute\_gradients} (Line~8 in
\autoref{list:api-example-1}), users can directly invoke
\texttt{Cluster.compute\_gradients}.

\PHM{Realizing \texttt{Worker} Declarations.}
For methods annotated with the \texttt{register} decorator, \texttt{Cluster}
enters the \texttt{execute\_all} path, which calls the target method on every
constituent \texttt{Worker} and aggregates results via \texttt{ray.get} (Lines~6--11).
For methods annotated with \texttt{hw\_mapping}, \texttt{Cluster} inspects the
\texttt{tag\_name} argument, filters for \texttt{Worker}s whose resource type matches the
preferred hardware, and routes the request to those \texttt{Worker}s (Lines~13--19).
For methods annotated with \texttt{register\_serverless}, \texttt{Cluster}
replaces the reward proxy attribute (\texttt{attr}) with a callable that invokes
the registered serverless URL, so reward computation is performed by the
serverless backend (Lines~21--25).
In all affinity-based paths, if the preferred target is temporarily
unavailable, \texttt{Cluster} redirects execution to a compatible fallback
provided by the resource manager, ensuring forward progress under transient
contention.

The abstractions and mechanisms above, including \texttt{Worker} logic, decorator
annotations, resource bindings, and \texttt{Cluster}-level
realization, constitute what users configure at pipeline setup time. Once the
pipeline is materialized, the control plane takes over: it drives
trajectory-level rollout, reward evaluation, and asynchronous training without
further user intervention, which we explain next.

\section{System-Managed Control Plane}
\label{sec:runtime}

\SystemName{}'s control plane is entirely \emph{system-managed}: users specify no
coordination logic. In this section, we first describe how the control
plane orchestrates trajectory-level rollout, reward, and asynchronous training
(\S\ref{subsec:async-workflow}--\S\ref{subsec:async-training}), then present
cross-cutting optimizations that span the data and control planes to improve
end-to-end efficiency (\S\ref{subsec:cross-cutting-opt}).

\begin{figure}
    \centering
    \includegraphics[width=\linewidth]{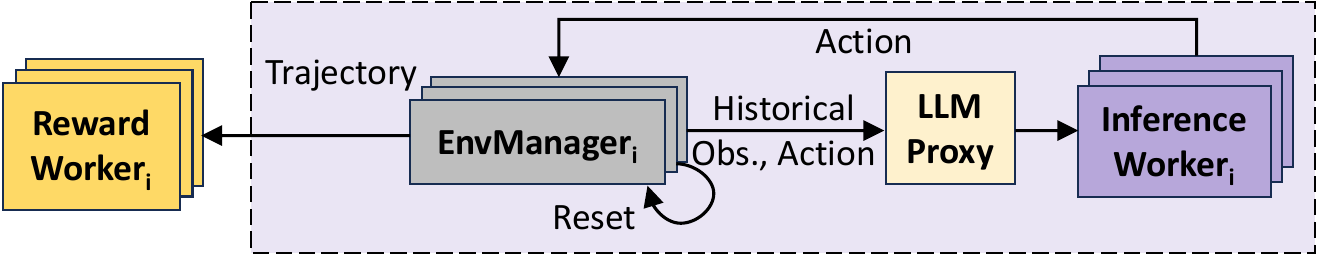}
    \vspace{-10pt}
    \caption{Trajectory-Level Rollout and Reward Overview.}
    \label{fig:traj_rollout}
\end{figure}

\subsection{Trajectory-Level Rollout and Reward}
\label{subsec:async-workflow}

As identified in \S\ref{sec:motivation} (\textbf{R2}), batched environment
interaction forces fast environments to wait for stragglers, inflating
end-to-end rollout latency. To eliminate this synchronization barrier,
\SystemName{} adopts a \emph{trajectory-level} design in which each trajectory
progresses through generation, environment interaction, and reward computation
independently. \autoref{fig:traj_rollout} gives an overview.
Three components realize this: \texttt{LLMProxy} dispatches
requests at per-trajectory granularity, \texttt{EnvManager} drives each
environment's lifecycle independently, and \texttt{Reward Worker}s score asynchronously.

\PHM{\texttt{LLMProxy}: Trajectory-Level LLM Generation.}
As shown in \autoref{fig:sys_arch}, \texttt{LLMProxy} sits between
\texttt{EnvManager}s and inference workers in the control plane.
It acts as a gateway that decouples generation clients from the
serving instances, dispatching per-trajectory requests across
inference workers.

Each inference worker runs a command-driven event loop
(\autoref{fig:traj_rollout}) that manages an inference engine (e.g.,
vLLM~\cite{vllm}, SGLang~\cite{sglang}). This loop operates continuously in a
non-blocking fashion with two components: 

\emph{\uline{1) Step Wise Command Processing.}} The loop polls for commands dispatched by \texttt{LLMProxy}, \texttt{ADD} to enqueue requests and \texttt{ABORT} to cancel existing ones. When no commands are pending, it advances the engine by executing a decode or prefill step for a batch of requests, keeping GPU utilization high. Because commands are processed between engine steps, adding or aborting a trajectory does not stall ongoing generation.

\emph{\uline{2) Post-Processing.}} When the engine finishes a request, the loop
immediately invokes a pre-registered callback that post-processes the output and
returns the result to the requesting \texttt{EnvManager}. This allows each
trajectory to proceed to environment interaction as soon as its generation
completes, without waiting for stragglers.

\PHM{\texttt{EnvManager}: Trajectory-level Environment Interaction.} Each
\texttt{EnvManager} is a lightweight controller that manages the lifecycle of a
single environment to collect trajectories (also depicted in
\autoref{fig:traj_rollout}). It begins with environment initialization via
\texttt{reset}, after which it enters an independent event loop that
orchestrates the interaction between an environment instance and the shared
\texttt{LLMProxy} via \texttt{step}. During this loop, the \texttt{EnvManager}
maintains a list of (observation, action) pairs to construct a trajectory.
Specifically, it feeds \texttt{LLMProxy} with the historical (observation,
action) sequence as input to obtain the next action, applies this action to the
environment via \texttt{step}, and records the resulting observation.
Unlike batched environment interaction
(\autoref{fig:sub-batched_env_interaction}), where all environments must
synchronize before the next generation step, each \texttt{EnvManager} operates
on its own timeline, so a slow environment never blocks others.

\PHM{Overlapping Rollout and Reward.} Once a trajectory completes, the runtime
dispatches it to a reward worker, which invokes a serverless reward
function as a non-blocking task. Because each \texttt{EnvManager} yields
trajectories independently, reward computation begins as soon as any single
trajectory finishes, without waiting for an entire batch. \SystemName{} launches
multiple inference workers, \texttt{EnvManager}s, and reward workers so
that generation, environment interaction, and reward computation overlap in
time, hiding reward latency behind ongoing rollouts and maximizing pipeline
throughput.
This design avoids colocating reward on the
inference cluster, which would force batched
computation. 
Recent RL post-training
systems~\cite{roll,rollpacker,MiMo,areal,StreamRL} also adopt asynchronous reward
computation for this reason.
\SystemName{} further eliminates the
underutilization of dedicated reward GPUs
(\autoref{fig:reward_utilization}) through
serverless deployment.


\subsection{Asynchronous Training Orchestration}
\label{subsec:async-training}

\begin{figure}
    \centering
    \includegraphics[width=0.9\linewidth]{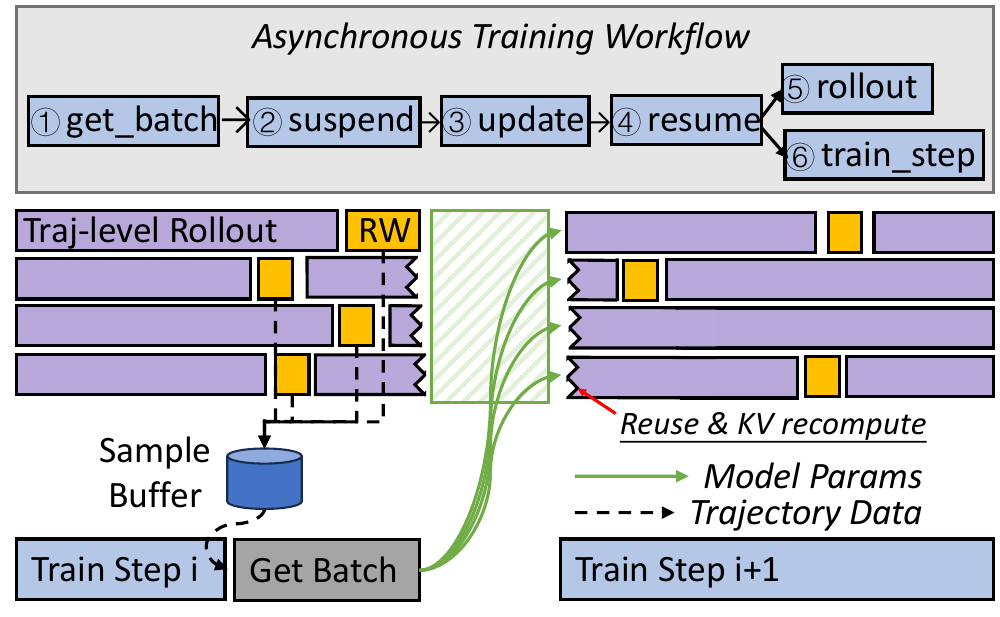}
    \caption{Asynchronous Training Workflow.}
    \label{fig:rollout_scheduler}
\end{figure}

In a disaggregated setup, synchronous training forces rollout workers to idle
while model weights are transferred across clusters.
This cost grows with model size (\S\ref{sec:motivation}, \textbf{R4}). 
To hide this overhead, \SystemName{} overlaps rollout and training: two stages execute concurrently on separate GPU clusters, coordinated through a weight
synchronization protocol (\autoref{fig:rollout_scheduler}).

\PHM{Weight Synchronization Protocol.} To propagate model weights to
inference workers while preserving the overlap between rollout and training,
\SystemName{} implements a six-step synchronization protocol in
each iteration.
\wcircle{1} \texttt{get\_batch}: The runtime invokes a blocking
\texttt{get\_batch} call to retrieve trajectories from the
\texttt{SampleBuffer}, which buffers scored trajectories for training
(\autoref{fig:sys_arch}). The pipeline waits until a predefined batch size is
collected.
\wcircle{2} \texttt{suspend}: Once the batch is ready, the runtime issues a
\texttt{suspend} command to the \texttt{LLMProxy}, preventing new generation
requests from being accepted while preserving in-flight trajectories.
\wcircle{3} \texttt{update}: After rollout is suspended, the pipeline fetches
the latest LLM weights from the training workers and updates all inference
workers.
\wcircle{4} \texttt{resume}: Upon completion, a \texttt{resume} command is
issued to the \texttt{LLMProxy}, which then continues pending generation
requests.
\wcircle{5} \texttt{recomp} \& \texttt{rollout}: To reuse in flight trajectories generated under the previous model version, \SystemName{} performs KV cache recomputation, rebuilding each trajectory's KV cache under the updated weights so that generation can continue without restarting from scratch.
\wcircle{6} \texttt{train\_step}: In parallel with the resumed rollout, the
pipeline executes \texttt{train\_step} on the batch retrieved in \wcircle{1},
allowing training to overlap with ongoing rollout.


\PHM{Asynchronous Bound and Buffer Management.} The protocol above lets rollout
and training proceed concurrently, but trajectories in \texttt{SampleBuffer} may
come from different model versions. Excessive staleness can increase variance
and destabilize training, so \SystemName{} enforces a per-trajectory
\emph{asynchronous bound} $\alpha$ (\textbf{R4}). If the current agent LLM is at
version $n$, any buffered trajectory must have been initiated by a version no
older than $(n - \alpha)$; trajectories outside this window are aborted.
This same version window bounds buffer growth: with $E$ concurrent
environments, \texttt{SampleBuffer} holds at most $O(\alpha \cdot E)$ pending
trajectories across versions. Before \texttt{get\_batch} forms a training batch,
it eagerly evicts stale trajectories, so highly asynchronous or out-of-order
completion cannot cause unbounded buffer growth.
Our empirical study (\S\ref{ssec:e2e-performance}) shows that $\alpha = 1$ balances training speed and stability.\footnote{Unlike AReaL~\cite{areal},
which bounds staleness only at trajectory start, \SystemName{} controls
trajectory-level staleness in each iteration to prevent long-tail trajectories
from spanning multiple model versions.}


\subsection{Cross-Cutting Optimizations}
\label{subsec:cross-cutting-opt}

The execution workflow above relies on data movement, hardware-specialized
serving, and scheduling mechanisms that span the data and control planes. We
describe three optimizations that improve end-to-end efficiency.

\PHM{Data Movement.} \SystemName{} matches each data path to an appropriate
transfer mechanism. 
The \textit{transfer protocol} streams trajectories and
supervision signals between stages as Ray's object references~\cite{ray}. 
The exchanged objects are sharded to match each worker's parallelism layout. The
\textit{model update group} synchronizes weights between stages. Intra-cluster
weight synchronization uses NCCL~\cite{nccl} over NVLink or InfiniBand. 
The main bottleneck is \emph{cross-cluster} weight updates, which bridge the training
cluster and inference cluster over lower-bandwidth
Ethernet. 
\SystemName{} addresses this with an \textit{asynchronous weight
update engine} built on Mooncake~\cite{qin2024mooncake}: after each training
step, updated weights are bucketized (e.g., 1GB) and published to a remote
CPU-resident Mooncake store rather than synchronously pushed to remote inference
workers. Inference workers then asynchronously fetch the latest weight buckets
from the Mooncake store on demand, decoupling weight transfer from ongoing
rollouts and avoiding GPU stalls caused by slow cross-cluster communication.

This makes the data movement explicit: training workers
write weight buckets once over the lower-bandwidth cross-cluster link, while
inference workers pull them over high-bandwidth intra-cluster links. 
\shepherd{The additional hop through the Mooncake store introduces only minimal overhead, as we quantify in \S\ref{subsec:cross-cutting-eval}. 
While Mooncake adds a pull step, it uses optimized data movement (e.g., zero-copy transfers) and is overlapped with ongoing rollout and weight push. }
For smaller models, the absolute transfer volume is also lower, limiting the exposed latency. 
We quantify both the accumulated pull cost and the non-overlapped exposed overhead in \S\ref{subsec:cross-cutting-eval}.

\PHM{Redundant Environment Rollouts.} Within the control plane, the
trajectory-level design of \texttt{LLMProxy} and \texttt{EnvManager} enables an
optimization we term \textit{redundant environment rollouts}. This
allows the system to launch more environments than required for trajectory
collection. Once the target number of trajectories has been collected, in-flight
trajectories can be terminated. Because rollouts are managed at trajectory
granularity, slow or failed environments do not block faster ones, mitigating
stragglers (\S\ref{subsec:cross-cutting-eval}) and environment failures
(\S\ref{sec:large-scale-analysis}).

\PHM{Prefill-Decoding (PD) Disaggregation.} Within the data plane, PD disaggregation, which is widely adopted in LLM serving systems~\cite{distserve,patel2023splitwise}, refines the \texttt{LLMProxy} routing described in \S\ref{subsec:async-workflow}. \SystemName{} supports PD disaggregation by
deploying prefill and decoding workers on compute-optimized and
bandwidth-optimized GPUs, respectively, and routing the two phases of each
request to the corresponding workers. 
We evaluate the above optimizations in \S\ref{subsec:cross-cutting-eval}.

\section{Evaluation}
\label{sec:eval}

In this section, we present the end-to-end evaluation
(\S\ref{ssec:e2e-performance}), ablations of the four design
requirements(\S\ref{subsec:eval-ablations}), cross-cutting optimizations
(\S\ref{subsec:cross-cutting-eval}), and the system's disaggregation tax
(\S\ref{subsec:eval-disaggregation-tax}).

\begin{figure*}[tb]
    \centering
    \begin{subfigure}[b]{0.422\linewidth}
        \centering
        \includegraphics[width=\linewidth]{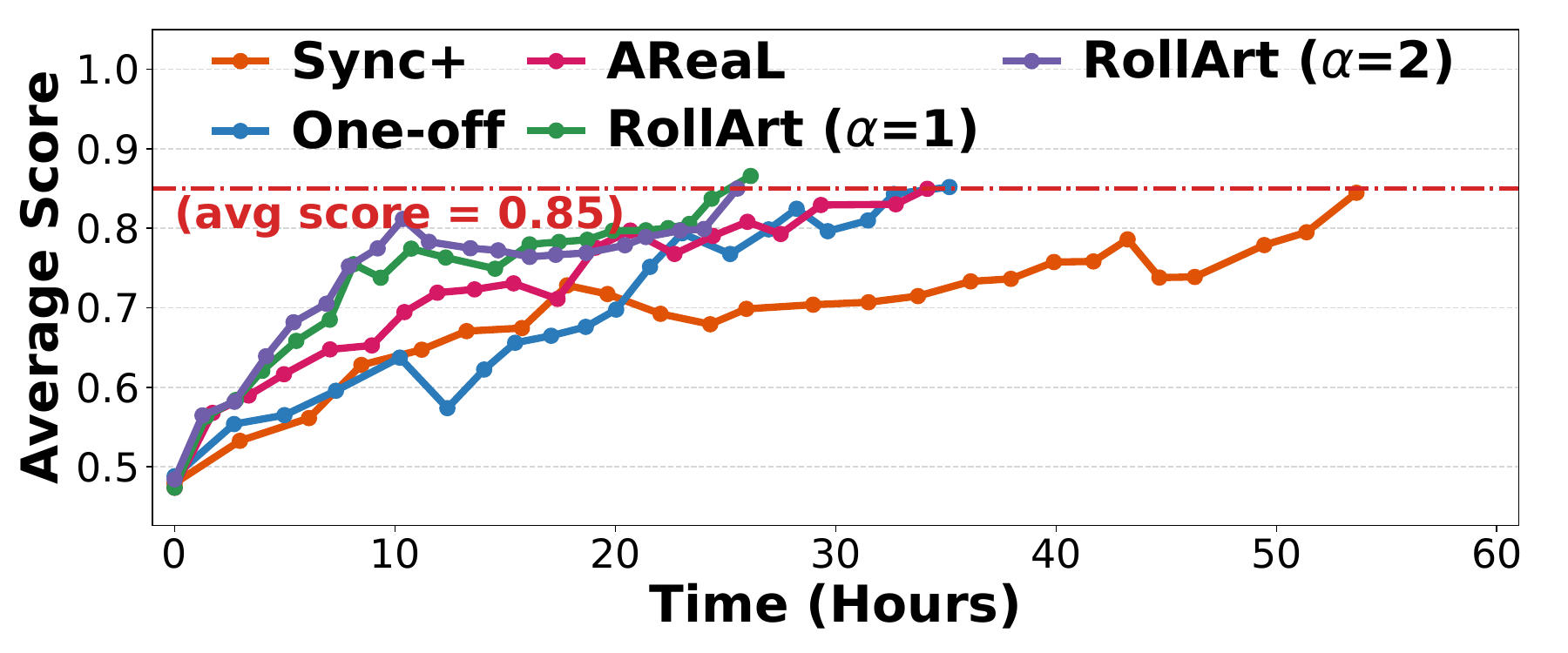}
        \caption{\shepherd{Time-to-Score (target $0.85$) on Qwen3-32B.}}
        \label{fig:e2e-acc}
    \end{subfigure}
    \begin{subfigure}[b]{0.28\linewidth}
        \centering
        \includegraphics[width=\linewidth]{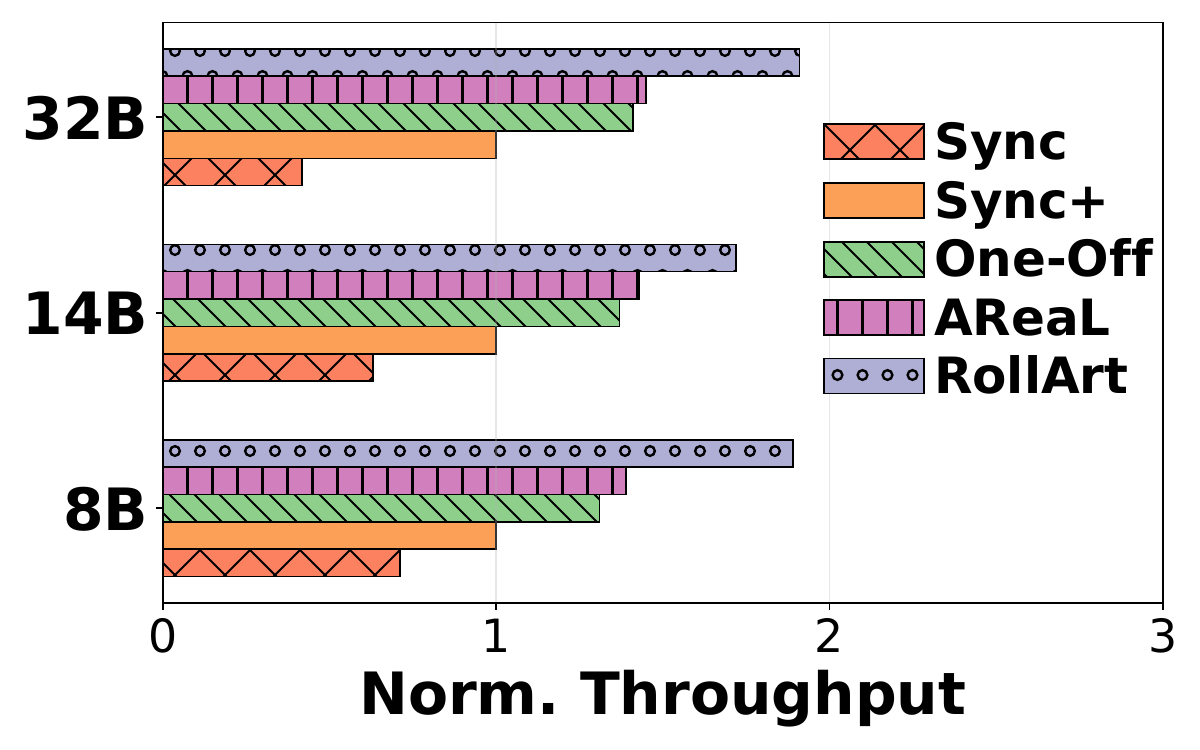}
        \caption{\shepherd{Norm. Throughput across LLMs.}}
        \label{fig:e2e-thr}
    \end{subfigure}
    \begin{subfigure}[b]{0.195\linewidth}
        \centering
        \includegraphics[width=\linewidth]{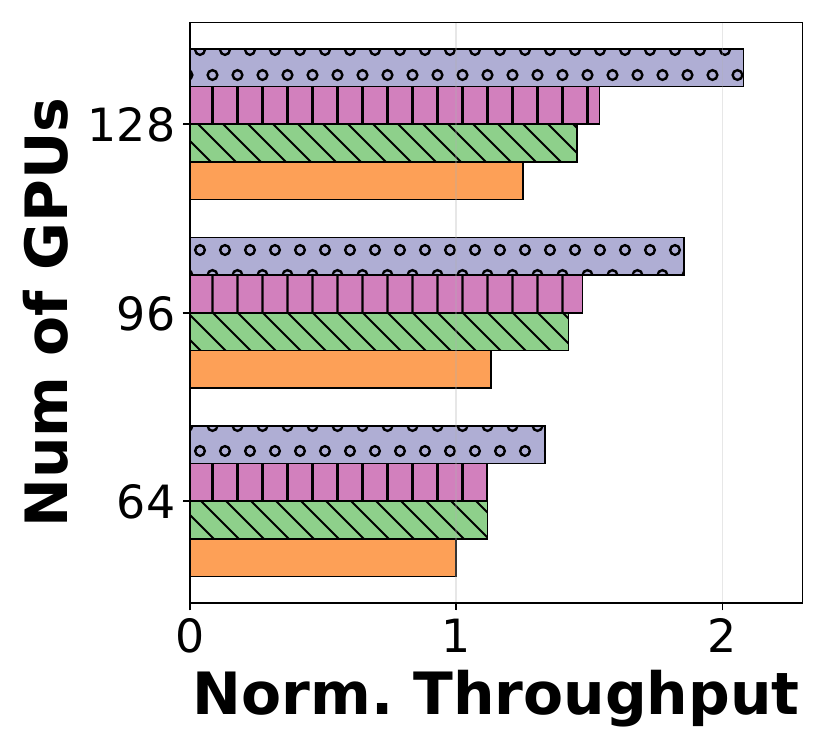}
        \caption{\shepherd{Scaling Efficiency.}}
        \label{fig:e2e-thr-scaling}
    \end{subfigure}
    \caption{End-to-end results: (a) time-to-score on Qwen3-32B; (b) throughput across LLMs 
    (normalized to Sync+); (c) throughput of Qwen3-14B across H800 GPU counts, normalized 
    to Sync+ on 64 H800 GPUs.}
    \label{fig:e2e-combined}
    \vspace{-13pt}
\end{figure*}

\subsection{Evaluation Setup}
\label{subsec:eval-setup}

\PHB{Models and Tasks.} We train Qwen3~\cite{yang2025qwen3technicalreport}
models from 8B to 32B on a diverse mix of agentic tasks
(\autoref{tab:env-taxomy}), using a 32k-token maximum context length. Because
\texttt{SWE-bench} is substantially harder than the other tasks, we train it
only on Qwen3-32B. For mathematical tasks, we use Qwen2.5-7B as the reward LLM
to validate the reasoning process.

\PHM{Infrastructure.} We deploy \SystemName{} on a 96-GPU H800 cluster and a 32-GPU
H20 cluster. GPU nodes within each cluster are connected by 400 Gbps
InfiniBand; cross-cluster communication uses 200 Gbps Ethernet. We use one
dedicated CPU cluster for \texttt{SWE-bench}, another for the remaining
environments, and our internal serverless platform for reward workers. Unless
otherwise stated, all experiments use 128 GPUs. We use NCCL~\cite{nccl} v2.26.5 for intra-cluster weight updates and a
Mooncake v0.3.7~\cite{qin2024mooncake} storage server for cross-cluster
communication.

\PHM{Training Configuration.} We train with GRPO~\cite{he2025deepmath}, a
batch size of 512, a group size of 8, and uniform task sampling. For
asynchronous training, we reserve 32 H800 GPUs for training and use the
remaining H20 and H800 GPUs for rollouts. Rollout tensor-parallelism degrees
for Qwen3-8B/14B/32B are $1$, $2$, and $4$, respectively; the corresponding
training-side tensor and pipeline parallelism are tuned for throughput. Rollouts run on
vLLM 0.8.4 with prefix caching and CUDA graphs enabled, while training runs on
Megatron v0.12.2.


\PHM{Baselines.} We compare against four post-training baselines,
grouped into sync and async designs. Since no existing open-source
system supports our full spectrum of agentic tasks, we implement two sync
baselines on top of \SystemName{}: \textbf{Sync}, a standard synchronous RL
pipeline, and \textbf{Sync+}, which strengthens Sync with async reward
computation, async environment interaction, and serverless
offloading---techniques that are widely adopted in recent
systems~\cite{areal,StreamRL,roll}. For async baselines, we compare with
\textbf{One-off}~\cite{deepscaler2025} (see
\autoref{fig:paradigm-all-in-one}-Right) and \textbf{AReaL}. One-off overlaps rollout and training by
consuming trajectories from the previous step, but requires all trajectories in
an iteration to finish with stale model weights. Differently, \SystemName{} with
$\alpha=1$ preempts and resumes ongoing rollouts in the next iteration. \areal{We re-implemented AReaL in \SystemName{}’s codebase, adopting AReaL’s staleness bound of one, which constrains trajectory-level staleness only at the start rather than at every turn of every trajectory, to ensure fairness.} Both One-off and AReaL implementations incorporate the
Sync+ optimizations, so they differ from \SystemName{} mainly in the coordination between training and rollout for staleness control, as well as the lack of hardware affinity routing. \textbf{Laminar}~\cite{Laminar}, another async baseline, repacks long-tail rollouts but does not account for hardware
affinity, serverless reward computation, or long-tail environment execution, and we provide a quantitative
feature-decomposition argument in \S\ref{ssec:e2e-performance} (Throughput Efficiency). We run baselines on 128 H800 GPUs,
so \SystemName{} incurs roughly $83\%$ of the baselines' per-GPU-hour cost.

\PHM{Metrics.} We measure end-to-end latency as the average step time over five
iterations, and throughput as the number of prompt and response tokens in a
global batch divided by step time~\cite{sheng2025hybridflow}. We report the
average validation score across tasks.

\subsection{End-to-End Evaluation}
\label{ssec:e2e-performance}

\PHB{Model Convergence.} We measure validation score every ten iterations on Qwen3-32B, the most network-sensitive model in our setup, and report the time to reach a target score of $0.85$ in \autoref{fig:e2e-acc}.
The asynchronous configuration with $\alpha=1$ reduces step time by
$2.05\times$, $1.35\times$, \areal{and $1.31\times$} over Sync+, One-off, and AReaL,
respectively. One-off already improves over Sync+ by $1.52\times$ by
overlapping rollout and training. AReaL converges faster in the early stage but achieves similar time-to-score performance to the One-off approach because it bounds staleness only at the start of trajectories, collecting more stale long-tail trajectories. \SystemName{} enforces per-step staleness control within the bound $\alpha$ in \S\ref{subsec:eval-ablations} and promptly aborts stale trajectories. Increasing
the bound to $\alpha=2$ for \SystemName{} improves initial convergence but slightly worsens later
time-to-score relative to $\alpha=1$, consistent with the throughput--quality
tradeoff in \S\ref{subsec:eval-ablations}. 
The results in \autoref{fig:e2e-thr} corroborate the advantages of \SystemName{} over these baselines across 8B, 14B, and 32B.

\PHM{Throughput Efficiency.} \autoref{fig:e2e-thr} reports throughput efficiency
normalized to Sync+; we set $\alpha=1$ for \SystemName{}. The results show how
throughput improves as the baselines add more overlap and scheduling
flexibility. Sync+ first improves throughput by
$1.40\text{--}2.40\times$ over Sync by adding async reward computation, async
environment interaction, and serverless offloading. One-off adds
$1.31\text{--}1.47\times$ over Sync+ by overlapping rollout and training.
\areal{AReaL contributes another
$1.03\text{--}1.06\times$ over One-off. \SystemName{} then adds
$1.22\text{--}1.36\times$ over AReaL by layering hardware-affinity mapping
on bounded-staleness async training. This \SystemName{}/AReaL gap is consistent
with the hardware-affinity contribution measured in isolation in
\S\ref{subsec:eval-ablations} (R1, $1.12\text{--}1.37\times$ over an H800-only
rollout).} Overall, \SystemName{} achieves $2.65\text{--}4.58\times$ throughput
over Sync.

Laminar~\cite{Laminar} is closed source, so we cannot evaluate it directly. Instead, the decomposition above isolates the capabilities that Laminar does not provide. The measured \SystemName{}/AReaL gap captures the effect of hardware affinity (\S\ref{subsec:eval-ablations}, R1). Laminar does not optimize for environment instability, which yields a $1.23\text{--}2.27\times$ improvement under environment variability (\S\ref{subsec:eval-ablations}, R2), and it lacks serverless reward computation, which can reduce rollout time by up to $\sim 2\times$ (\S\ref{subsec:eval-ablations}, R3). Composing these isolated gains lower bounds the gap between \SystemName{} and Laminar. 


\PHM{Scaling Efficiency.} We train Qwen3-14B while sweeping the cluster size from 64 to 128 H800 GPUs. Thus, \SystemName{} does not support hardware-affinity mapping in this evaluation. We compare the scaling performance of \SystemName{} under $\alpha=1$. \autoref{fig:e2e-thr-scaling} reports throughput normalized to Sync+ on 64 GPUs. \areal{As cluster size increases, Sync+ exhibits diminishing marginal gains, and One-off and AReaL show similar limited scaling performance from 96 to 128 GPUs, whereas \SystemName{} delivers $1.33\text{--}2.08\times$ higher throughput, demonstrating better scaling efficiency.}

\subsection{Ablations of Design Requirements}
\label{subsec:eval-ablations}

We next present ablations of the four design requirements.

\begin{figure}[tb]
    \centering
    \begin{subfigure}{0.47\linewidth}
        \centering
        \includegraphics[width=\linewidth]{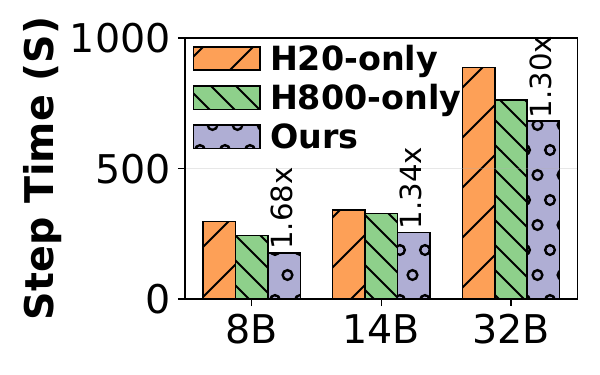}
        \caption{Rollout Efficiency.}
        \label{fig:compute-efficiency}
    \end{subfigure}
    \hfill
    \begin{subfigure}{0.23\textwidth}
        \centering
        \includegraphics[width=\linewidth]{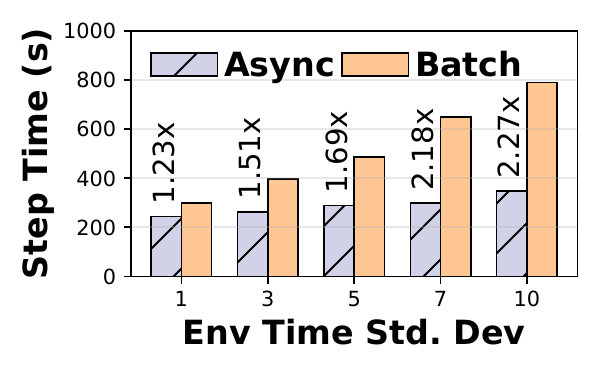}
        \caption{Traj.-Level vs. Batch Rollout.}
        \label{subfig:async_env_batch}
    \end{subfigure}
    \caption{The efficiency of (a) hardware affinity and (b) trajectory-level rollout, where 
    env latency is sampled from a Gaussian distribution with mean $\mu = 10$~s and standard deviation $\sigma$ 
    varying from 1 to 10~s (x-axis).}
    \label{fig:rollout_col_disagg}
\end{figure}

\PHM{R1: Hardware-Affinity Mapping.}
We evaluate hardware-affinity mapping across LLM sizes on compute-optimized and
bandwidth-optimized GPUs. To isolate routing, we fix training to 32 H800 GPUs
and compare three rollout configurations that yield approximately an equal cost: 72 H800
GPUs, 208 H20 GPUs, and an affinity-aware mix of 64 H800 plus 24 H20 GPUs. The mixed
configuration routes mathematical and game-oriented agentic tasks to H20 GPUs
while keeping prefill-heavy work on H800 GPUs. As shown in
\autoref{fig:compute-efficiency}, \SystemName{} reduces step time by
$1.30\text{--}1.68\times$ over H20-only and by $1.12\text{--}1.37\times$ over
H800-only. H20-only performs worst because many agentic tasks still incur
frequent prefill operations, while the mixed configuration adds H20 capacity for
decoding-heavy tasks without moving prefill-heavy work off compute-optimized
GPUs. Beyond task-level mapping, PD disaggregation
extends \SystemName{}'s affinity routing from task-level to phase-level
placement; we evaluate it as a cross-cutting optimization in
\S\ref{subsec:cross-cutting-eval}.



\PHM{R2: Trajectory-Level Asynchrony.}
Trajectory-level rollout decouples environment interaction across trajectories,
so slow turns do not stall an entire batch (\S\ref{subsec:async-workflow}). To
isolate this effect, we run Qwen3-8B with a 32k context and inject per-turn
environment latency sampled from Gaussian distributions with mean $\mu = 10$~s
and standard deviation $\sigma$ ranging from $1$~s to $10$~s. This synthetic
injection controls latency variance; production-side straggler stacking is
reported in \S\ref{sec:large-scale-analysis}. \autoref{subfig:async_env_batch}
compares trajectory-level and batch-level environment interaction using average
step time over ten iterations. As $\sigma$ increases, trajectory-level rollout
improves over batch-level interaction from $1.23\times$ to $2.27\times$, showing
that it better absorbs environment variance.

\begin{figure}[tb]
    \centering
    \includegraphics[width=\linewidth]{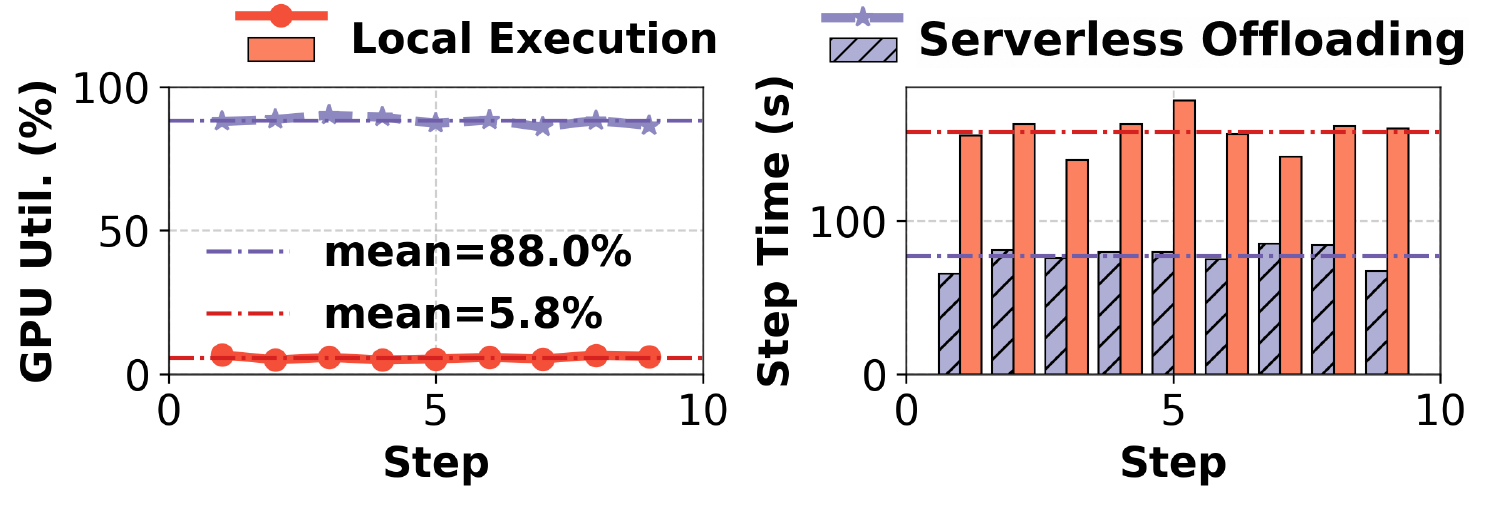}
    \vspace{-15pt}
    \caption{Comparison between dedicating local GPUs and using serverless offloading.}
    \label{fig:reward_comparison_exp}
    \vspace{-5pt}
\end{figure}

\PHM{R3: Serverless Offloading.}
We evaluate serverless offloading against a local-GPU reward setup on a 16-H800
cluster. We run three concurrent mathematical agentic RL jobs with Qwen3-8B/16k
as the actor agent and Qwen2.5-7B as the reward LLM. Both serverless and local
setups reserve eight GPUs for training. The local setup assigns four remaining
GPUs to rollout and the other four to reward, whereas the serverless setup
assigns all eight remaining GPUs to rollout and offloads reward computation to
an elastic serverless platform. \autoref{fig:reward_comparison_exp} reports GPU
utilization and per-step rollout time for a batch size of 84, including
asynchronous reward computation. Serverless offloading raises average GPU
utilization from $6\%$ to $88\%$ and doubles the local rollout allocation,
reducing average rollout time from $158$~s to $77$~s. Although remote
serverless reward calls introduce network I/O, the disaggregation tax
measured in \S\ref{subsec:eval-disaggregation-tax} is small (max
$2.1$~s, mean $0.01$~s per call). Thus, transfer cost does not negate the resource efficiency gains from serverless offloading.


\begin{figure}[tb]
    \centering
    \includegraphics[width=0.9\linewidth]{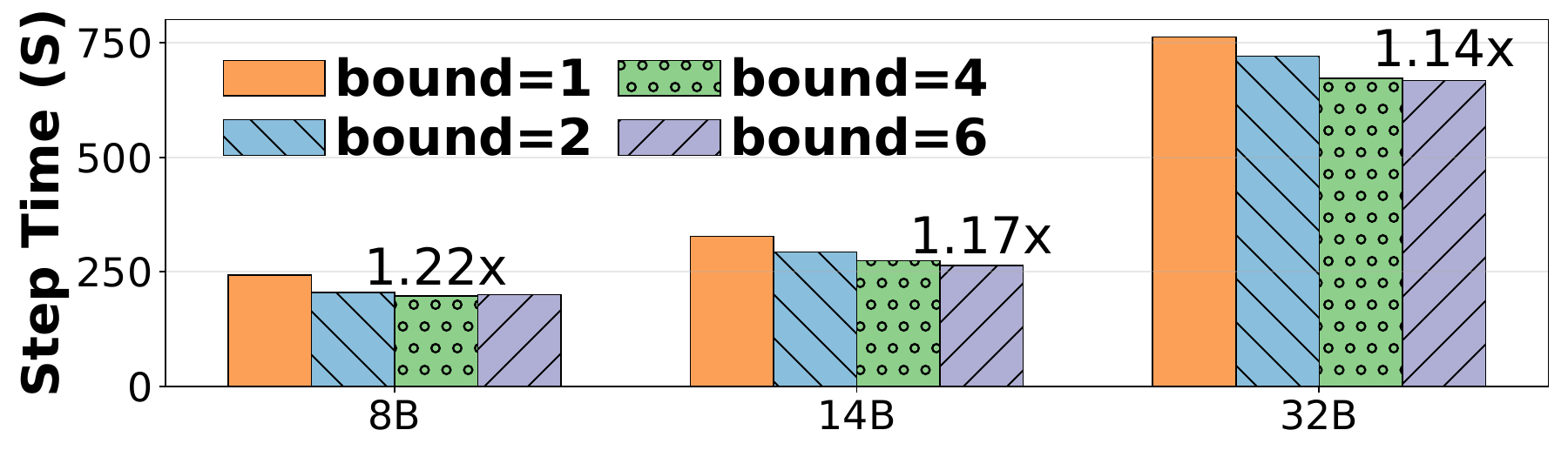}
    \vspace{-5pt}
    \caption{Average step time across LLMs and bounds.}
    \label{fig:async-ratio}
\end{figure}

\PHM{R4: Asynchronous Bound.}
We quantify the throughput--convergence tradeoff induced by the per-trajectory
asynchronous bound $\alpha$. \autoref{fig:async-ratio} sweeps $\alpha$ from $1$
to $6$ and reports average step time across LLMs. Larger bounds reduce
staleness-triggered trajectory aborts and lower step time, but the
gain quickly plateaus: the best bound differs by LLM and improves step time by
at most $1.22\times$ over $\alpha=1$. These throughput gains do not
necessarily improve time-to-score. 
\autoref{fig:e2e-acc} shows that $\alpha=2$
already incurs a measurable late-stage time-to-score regression
, and we set $\alpha=1$ by default.


\begin{figure}[tb]
    \centering
    \begin{subfigure}{0.47\linewidth}
        \centering
        \includegraphics[width=\linewidth]{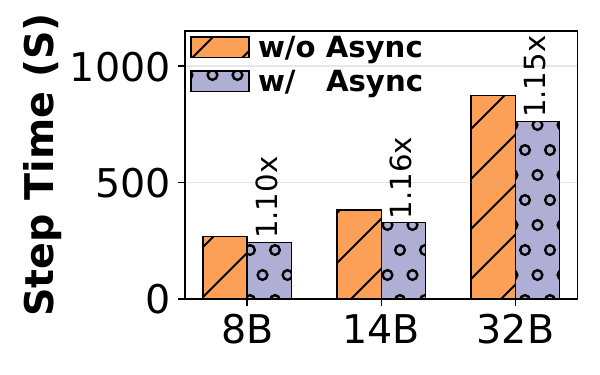}
        \caption{Cross-Cluster Comm.}
        \label{fig:comm-efficiency}
    \end{subfigure}
    \begin{subfigure}{0.23\textwidth}
        \centering
        \includegraphics[width=\linewidth]{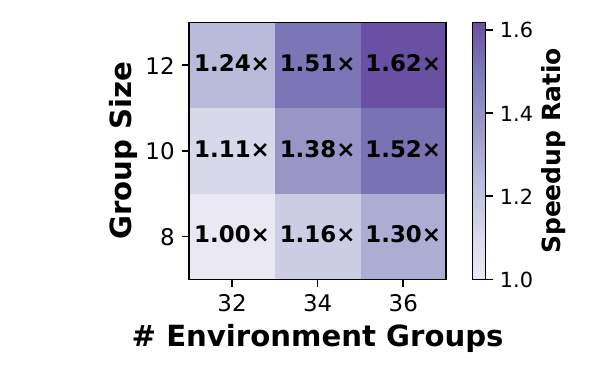}
        \caption{Redundant Env Rollout.}
        \label{subfig:env-red-rollout}
    \end{subfigure}
    \vspace{-5pt}
    \caption{Cross-cutting optimizations: (a) async cross-cluster weight 
    transfer; (b) redundant environment rollouts.}
    \label{fig:rollout_env_async}
\end{figure}

\subsection{Cross-Cutting Optimizations}
\label{subsec:cross-cutting-eval}

We evaluate the three cross-cutting optimizations in
\S\ref{subsec:cross-cutting-opt}.

\PHM{Asynchronous Weight Transfer.} Cross-cluster weight transfer is costly when
training and rollout clusters communicate over heterogeneous interconnects. To
quantify the benefit of overlap, \autoref{fig:comm-efficiency} compares
\SystemName{}'s asynchronous cross-cluster communication with veRL's NCCL-based
scheme, which assumes uniformly high-bandwidth links across servers.
Asynchronous communication reduces end-to-end step time by
$1.10\text{--}1.16\times$ across LLMs.
\autoref{tab:weight-sync-overhead} decomposes this gain. \emph{Weight
Push} is the cost of streaming updated weights from the training cluster to the
Mooncake store over the cross-cluster TCP/Ethernet link. \emph{Accumulated
Weight Pull} is the total cost of inference workers pulling new weight buckets
from Mooncake, while \emph{Exposed Weight Pull} is the residual pull cost not
hidden by ongoing rollout. The \emph{Naive (Push + Acc.\ Pull)} row reports the
cost exposed by a synchronous design such as veRL. Asynchronous overlap hides
$67\text{--}78\%$ of the pull cost, reducing the exposed overhead to at most
$9.6$~s on 32B versus $38.6\text{--}157.0$~s without overlap.

\begin{table}[t]
\centering
\caption{Breakdown of \SystemName{}'s asynchronous cross-cluster
weight transfer (seconds).}
\label{tab:weight-sync-overhead}
\small
\begin{tabular}{l|ccc}
\hline
\textbf{Cost} & \textbf{8B} & \textbf{14B} & \textbf{32B} \\
\hline
Naive (Push + Acc.\ Pull, no overlap) & 38.6 & 84.1 & 157.0 \\
\hline
Weight Push                          & 32.4 & 67.8 & 127.3 \\
Accumulated Weight Pull              & 6.2  & 16.3 & 29.7  \\
Exposed Weight Pull (\SystemName{})  & 1.4  & 5.1  & 9.6   \\
\hline
\end{tabular}
\end{table}

\PHM{Redundant Environment Rollouts.} Redundant rollouts exploit GRPO's group
structure to mask environment failures and stragglers. We run Qwen3-8B/32k on
32 H800 GPUs to vary the number of environment groups and the group size
on the \texttt{GEM-math} task. \autoref{subfig:env-red-rollout} reports rollout
speedup ratios. Larger groups and more environment groups increase
the chance that a fast subset finishes early, sustaining efficiency under
straggler conditions. 
The maximum speedup reaches $1.62\times$, and increasing
either parameter yields a speedup.


\PHM{PD Disaggregation.} PD disaggregation extends hardware-affinity
mapping from task-level to phase-level placement. 
\shepherd{
We train Qwen3-32B and Qwen3-30B-A3B models on \texttt{SWE} tasks with batch size of 128 and 32k sequence length.
We evaluate two configurations: 1P3D and 2P2D where each prefill node uses eight H800 GPUs and each decoding node uses eight H20 GPUs.
For the dense model, PD disaggregation achieves relatively modest speedups of 1.03$\times$ and 1.05$\times$ compared with colocation.
For the MoE model, 1P3D and 2P2D achieve $1.11\times$ and $1.21\times$ speedup over colocation, respectively.
The performance and optimal configuration\footnote{We also evaluated the 3P1D configuration, and both models achieved worst rollout performance, as the single decoding node became a bottleneck.} of PD disaggregation are workload- and model-dependent. \S\ref{sec:discussion} further discusses how to automate these configurations to adapt to workload demands. 
}


\subsection{Disaggregation Tax}
\label{subsec:eval-disaggregation-tax}

We next quantify the disaggregation tax
along three data paths. 


\PHM{Cross-Cluster Weight Synchronization.} Weight transfer is the
largest overhead, but the asynchronous data-movement in \S\ref{subsec:cross-cutting-eval} hides most of it. As
\autoref{tab:weight-sync-overhead} shows, \SystemName{} exposes $1.4\text{--}9.6$~s of residual pull cost; without overlap, rollout would block on Push + Accumulated Pull for $38.6\text{--}157.0$~s.

\PHM{Env-Interaction I/O.} Trajectory-level rollout moves data
between the environment and inference clusters at each interaction. Across our
evaluation, the maximum transferred volume is $2.7$~MB, and the maximum and
average per-call transfer overheads are $1.4$~s and $0.02$~s, respectively.
These costs are negligible relative to per-step training time.

\PHM{Serverless Reward I/O.} Remote reward calls transfer
trajectory payloads of up to $5.2$~MB, with maximum and average per-call
overheads of $2.1$~s and $0.01$~s. The residual disaggregation tax is therefore
dominated by cross-cluster weight transfer, which \SystemName{} overlaps with
rollout, so data movement does not erase the end-to-end gains reported in
\S\ref{ssec:e2e-performance}.



\begin{table}[t]
    \centering
    \caption{
        \shepherd{Rollout time (seconds) comparison between colocation and disagg. under different models and configurations.}
    }
    \label{tab:pd-disaggregation}
    \small
    \begin{tabular}{l|cc|cc}
    \hline
    \textbf{Model} & \textbf{1P3D} & \textbf{Colocate} & \textbf{2P2D} & \textbf{Colocate} \\
    \hline
    Qwen3-32B & 722.7 & 741.2 & 701.6 & 734.9 \\
    Qwen3-30B-A3B & 294.8 & 327.4 & 251.1 & 305.2 \\
    \hline
    \end{tabular}
\end{table}

\begin{figure*}[tb]
    \centering
    \begin{subfigure}[b]{0.33\linewidth}
        \centering
        \includegraphics[width=\linewidth]{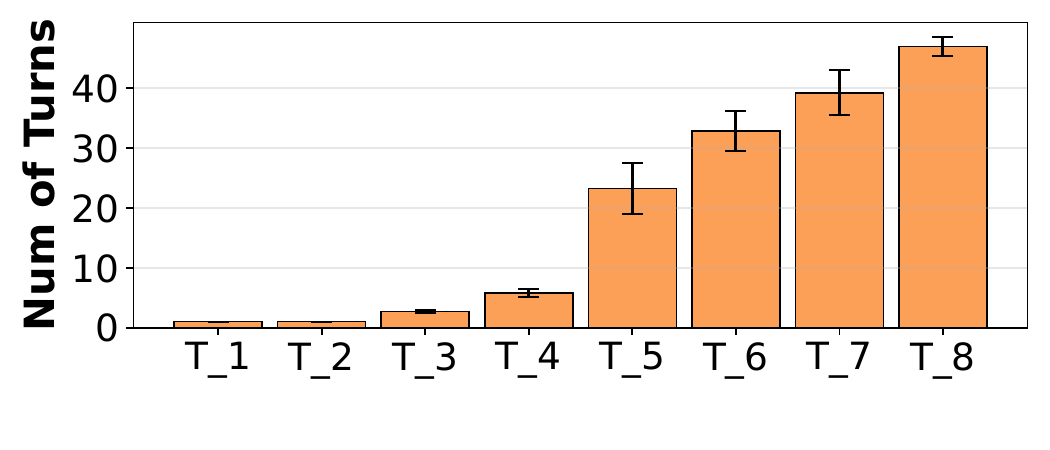}
        \caption{Average Number of Turns per Task.}
        \label{fig:action}
    \end{subfigure}
    \begin{subfigure}[b]{0.33\linewidth}
        \centering
        \includegraphics[width=\linewidth]{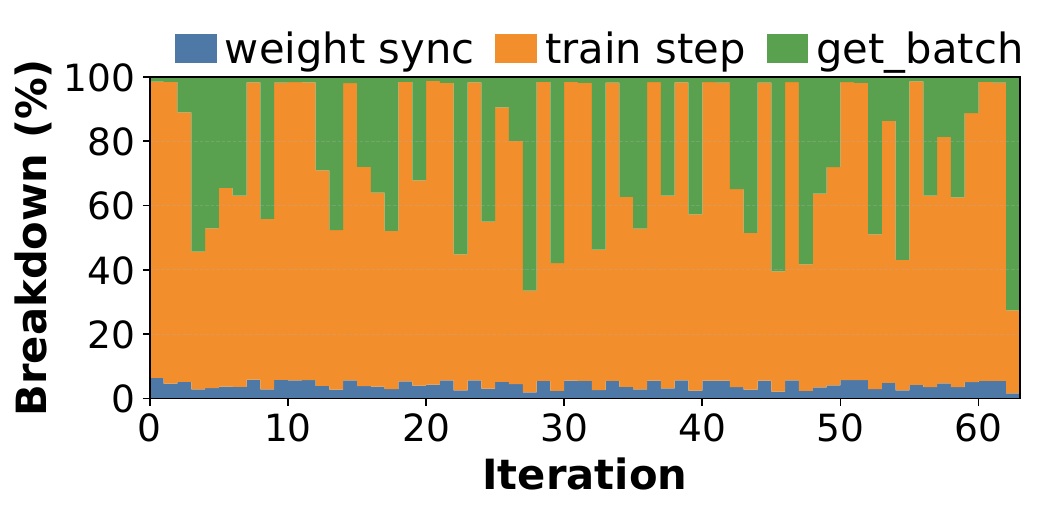}
        \caption{Step Time Distribution.}
        \label{fig:trace_dist}
    \end{subfigure}
    \begin{subfigure}[b]{0.33\linewidth}
        \centering
        \includegraphics[width=\linewidth]{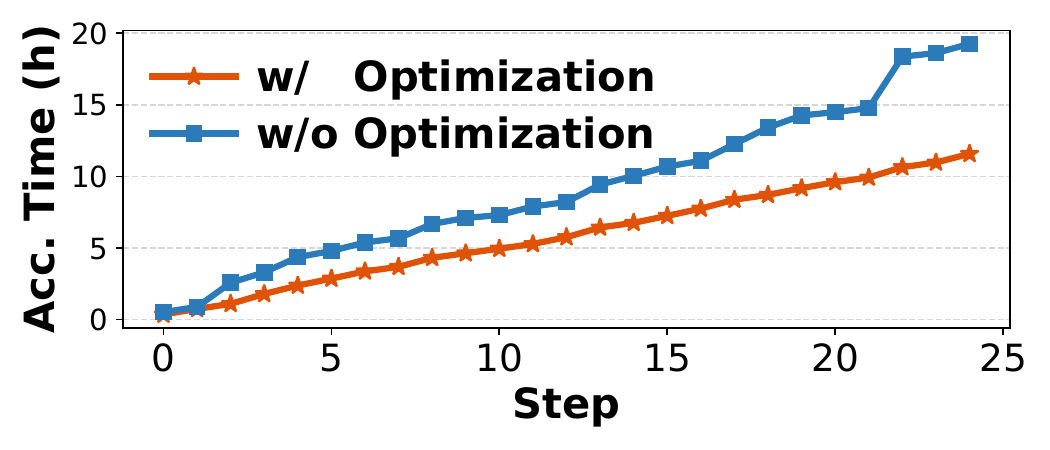}
        \caption{Accumulated Step Time.}
        \label{fig:opt_step_time}
    \end{subfigure}
    \caption{Production-grade agentic RL workload characterization and optimization.}
    \vspace{-15pt}
    \label{fig:workload_overview}
\end{figure*}

\section{\SystemName{} in Production}
\label{sec:large-scale-analysis}

Over the past nine months, thousands of agentic RL jobs have used \SystemName{}
for post-training. 
We report one large production deployment on more than 3000
GPUs to show how the workload properties in \S\ref{sec:motivation} appear in
practice and how operators tune \SystemName{} at scale.

\PHM{Workload Characterization.} We trained a hundreds-of-billions-parameter
MoE LLM with \SystemName{} on in-house mathematical and software-engineering
agentic tasks. Prompts and responses reach 12k and 46k tokens, respectively, and
the average number of turns per task ranges from 1 to 48
(\autoref{fig:action}). This mix exposes both prefill-heavy and decode-heavy
behavior. Stragglers persist throughout training: in each step, the
maximum response length exceeds 5$\times$ the mean and peaks at 9$\times$; the
maximum number of environment-interaction turns remains above 40$\times$ the
mean. These tails explain why production runs need both fast prefill execution
and trajectory-level rollout that can absorb slow environments.

The job uses asynchronous training with a 1:5 ratio of training to generation
GPUs and an asynchronous bound of one to balance rollout throughput with
gradient stability. Still, the longest iteration reaches 1.5 hours
(\autoref{fig:trace_dist}). The bottleneck is the blocking
\texttt{get\_batch} call: after computing log probabilities and gradients, the
training stage waits for \texttt{SampleBuffer} to collect enough trajectories.
This accounts for up to 62\% of iteration time as GPU idleness; removing it
would ideally reduce training time by 22\%
(\autoref{fig:trace_dist}).

\PHM{Characterization-Driven Optimization.} The production trace also shows why
coarse-grained characterization is a practical starting point for hardware
affinity: task domains are known before training, and their token and turn
profiles differ substantially. Using this information, we adjust the
training-to-generation resource ratio and tune prefix caching for the MoE
architecture. \autoref{fig:opt_step_time} shows that these changes achieve a
1.66$\times$ end-to-end speedup over the first 25 steps. Beyond this resource
tuning, we apply dedicated optimizations for environment stability and failure
recovery.

\PHM{Optimizing Environment Stability.} Across 60 iterations, long environment initialization failures appear in seven iterations. At this scale,
Docker-based environments on Kubernetes are sensitive to image-pull failures and
network instability during \texttt{env.reset}. We therefore use a multi-tiered
cache: an internal image registry mirrors external images, and a distributed
load-balanced cache between compute nodes and the registry absorbs high-volume
requests. During \texttt{env.reset}, clients fetch images from cache first.
This optimization raises the \texttt{env.reset} success rate above 99.99\%.
It also avoids repeated image-pull retries, keeping over 99.99\% of initialization
under one minute. After the optimization, no more
than ten heavy-tailed initialization events occur among hundreds of thousands of
resets; trajectory-level rollout and redundant environments absorb the remaining
tails.


\PHM{System Resilience.} \SystemName{} isolates failures across environment,
reward, inference, and training workers. Persistent sessions and exponential
backoff reduce environment-connection timeouts; Kubernetes manages environment
workers, and the serverless platform manages reward workers. If an inference
worker fails, \SystemName{} first restarts it on the same GPU; after repeated
failure, the worker is removed and its stored trajectories resume on healthy
workers. Training-worker failures restart from the latest checkpoint. In a
week-long training run, we observed only one failure.

\section{Discussion and Related Work}
\label{sec:discussion}

\PHB{Automating Hardware-Affinity Mapping.} \SystemName{}
currently requires static, per-task-domain affinity declarations, which is a
real limitation when computation profiles shift mid-run. 
In our production deployment, however, per-domain profiles are stable: turn counts vary widely \emph{across} families but stay bounded \emph{within} each domain across iterations, so coarse \texttt{hw\_mapping} declarations needed no re-tuning across the week-long 3,000-GPU run.
The natural
extension is an online profiler integrated with the resource manager:
per-domain prefill/decode latency would let \SystemName{} re-route requests
when within-domain shifts occur (e.g., a domain alternating between long-observation
and long-response prompts) and the same profiling can drive
PD-disaggregation ratios (\S\ref{subsec:cross-cutting-opt}). Per-iteration optimality is not required: workload
profile is dominated by task-domain identity rather than per-step policy
variance, so profiling decisions stabilize over a few iterations.

\PHM{Operational Regime.} \SystemName{}
targets multi-task agentic RL on heterogeneous, disaggregated infrastructure;
it offers limited benefit on homogeneous clusters, where \textbf{R1}'s affinity routing
collapses to a no-op; on strict on-policy or compute-light RL, where \textbf{R4}'s
bounded-staleness overlap yields little; and on deployments without elastic
compute, where \textbf{R3}'s serverless offloading is unrealizable. Trajectory-level
rollout (\textbf{R2}) and redundant environment rollouts
(\S\ref{subsec:cross-cutting-eval}) still apply to any agentic RL workload
exposed to long-tail environments.


\PHM{RL Post-Training Systems.}
Many systems address the systems challenges of RL post-training. Early
frameworks~\cite{Harper_NeMo_a_toolkit,deepspeedchat,hu2024openrlhf,verl} adopt appropriate stage and resource mapping to improve utilization. Subsequent
systems~\cite{liu2025specrlacceleratingonpolicyreinforcement,chen2025respecoptimizingspeculativedecoding,cheng2025fastllmposttrainingdecoupled,shao2025beatlongtaildistributionaware,RhymeRL,zhong2024rlhfuseefficientrlhftraining,areal,asyncflow,Laminar,StreamRL}
explore a range of accelerations, including speculative decoding,
fusion of pipeline stages, asynchronous or
decoupled training to hide latency, and various forms of resource
disaggregation. \SystemName{} builds on the disaggregated
paradigm and assigns workloads based on hardware affinity.



\PHM{Resource Disaggregation.} Resource disaggregation is adopted in many modern systems to improve utilization~\cite{disaggregated-mem-isca09,shan2018legoos,guo2023mira}. This pattern is prevalent in LLM serving, where systems separate the prefill and decoding phases~\cite{distserve,
patel2023splitwise, hu2024inference, strati2024dejavu, qin2024mooncake,megascale-infer,step3}. Several RL
systems apply a similar principle, separating the training and rollout stages
across different GPU types~\cite{StreamRL,seamlessflow}. Compared to these
approaches, \SystemName{} provides a more general and fine-grained
disaggregation model, tailored for the entire lifecycle of multi-task agentic RL.

\section{Conclusion}
\label{sec:conclusion}

We presented \SystemName{}, a system for multi-task agentic RL training on
disaggregated infrastructure. Its design rests on four requirements:
hardware-affinity workload mapping (\textbf{R1}), trajectory-level asynchronous rollout
(\textbf{R2}), serverless offloading for stateless reward (\textbf{R3}), and bounded-staleness
asynchronous training (\textbf{R4}). On Qwen3 (8B--32B), \SystemName{} reduces step
time by $1.31\text{--}2.05\times$ over strengthened
synchronous, one-off, and AReaL baselines. A week-long, 3,000-GPU production run training
a hundreds-of-billions-parameter MoE model confirms these gains at scale.

\section*{Acknowledgment}
We thank our shepherd, Shadi Noghabi, and the anonymous reviewers for their 
valuable comments that help improve the quality of this work. 
This work was supported in part by the HKUST-Alibaba Joint Laboratory on Big Data and AI, RGC CRF Grant (Ref. \#C6015-23G), RGC\ GRF Grant (Ref. \#16217124), and NSFC/RGC CRS Grant (Ref. \#CRS\_HKUST601/24).


\bibliographystyle{plain}
\bibliography{reference}
\end{document}